\documentstyle[aps,preprint]{revtex}

\begin{document}
\draft
\tightenlines
\title
{\bf Models of the Pseudogap State of Two-Dimensional Systems}
\author{E.Z.Kuchinskii,\ M.V.Sadovskii}
\address
{Institute for Electrophysics, \\
Russian Academy of Sciences, Ural Branch, \\
Ekaterinburg 620049, Russia\\
E-mail: kuchinsk@ief.uran.ru,\ sadovski@ief.uran.ru}
\maketitle


\begin{abstract}
We analyze a number of ``nearly exactly'' solvable models of electronic
spectrum of two-dimensional systems with well-developed fluctuations of
short range order of ``dielectric'' (e.g. antiferromagnetic) or
``superconducting'' type, which lead to the formation of anisotropic
pseudogap state on certain parts of the Fermi surface. We formulate a
recurrence procedure to calculate one-electron Green's function which takes
into account all Feynman diagrams in perturbation series and is based upon
the approximate Ansatz for higher-order terms in this series.
Detailed results for spectral densities and density of states are presented.
We also discuss some important points concerning the justification of our
Ansatz for higher-order contributions.
\end{abstract}


\newpage
\tableofcontents

\newpage
\narrowtext

\section{Introduction.}

Recently there was an upsurge of interest in observation of the pseudogap in
the spectrum of elementary excitations of underdoped high-temperature 
superconductors (HTSC) \cite{R,RC}. These anomalies were observed in a number
of experiments, such as the measurements of optical conductivity, NMR,
inelastic neutron scattering and angle-resolved photoemission (ARPES)
(cf. review in \cite{R}). Probably most striking evidence for the existence
of this unusual state were obtained in ARPES--experiments\cite{Ding,D}, 
which demonstrated essentially anisotropic changes in the spectral density of
current carriers in rather wide temperature region in normal 
(nonsuperconducting) phase of these systems (cf. review in \cite{RC}). 
A remarkable anomaly observed in these experiments was that the most
significant changes (in comparison with the usual Fermi-liquid behavior)
in spectral density were seen close to the point $(\pi,0)$
in the Brillouin zone, while there were no changes at all in the direction of
zone diagonal (point $(\pi,\pi)$), which in fact demonstrates the 
``destruction'' of the Fermi surface around the point $(\pi,0)$ and 
conservation of Fermi--liquid behavior in the direction of zone diagonal.
In this sense it is usually claimed that the pseudogap symmetry is of the
``$d$-wave'' type, the same as the symmetry of superconducting energy gap
in these compounds \cite{R,RC}. At the same time the mere fact that these 
anomalies exist at the temperatures much higher than superconducting 
transition temperature, as well as in the underdoped (non optimal) region
of carrier concentrations, may signify some other nature of these anomalies,
not connected directly with Cooper pairing. Typical phase diagram of 
HTSC--system is shown in Fig.1.

There are rather many theoretical papers now attempting to interpret the
observed anomalies. We may classify these attempts in two main schools of
thought (scenarios). One is based upon the idea of Cooper pairs (fluctuation) 
formation at temperatures higher than the usual superconducting transition 
temperature\cite{R,Gesh,EK,Levin}. Another assumes that the pseudogap phenomena 
are determined mainly by fluctuations of antiferromagnetic (AFM) short range
order\cite{Kam,Bar,Pines,Sch,SchP}.

Rather long ago one of the authors of the present paper had proposed an
exactly solvable model of the pseudogap formation in one-dimensional system
due to well developed fluctuations of short range order of charge density
(CDW) or spin density (SDW) wave \cite{C1,C2,C79,C91,S91}.
Recently this model has attracted some interest in connection with attempts
to understand the pseudogap state in HTSC-cuprates\cite{Sch,SchP,McK,Tch1,Ren}.
In particular in Refs.\cite{Sch,SchP} an important generalization of this 
model was formulated for the case of two-dimensional electronic system with
random field of spin fluctuations (AFM short range order). In this model of
``hot spots'' on the Fermi surface \cite{Sch,SchP}, using the formal scheme
of Refs.\cite{C79,C91,S91}, they obtained rather detailed description of
pseudogap anomalies for the case of large enough temperatures (``weak 
pseudogap'' region in Fig.1). In Refs.\cite{Tch1,Ren} a simplified variant of
this model \cite{C1,C2}, corresponding to the limit of very large correlation
lengths of fluctuations of short range order, was used to describe the 
pseudogap state determined by well developed fluctuations of superconducting
(SC) short range order. In a recent paper \cite{PS} this (over) simplified 
model was used to derive Ginzburg--Landau expansion (for different types of 
Cooper pairing) for the system with strong fluctuations of CDW(SDW,AFM)--type, 
in the framework of ``hot patches'' model proposed in this work.
At the same time, in Ref.\cite{Tch2}, dedicated to rather detailed review of
the model proposed in Refs.\cite{C1,C2,C79,C91,S91}, a certain error was noted
in early papers\cite{C79,C91,S91} in the analysis of the case of finite 
correlation lengths of fluctuations of short range order. In Ref.\cite{SchP} 
it was claimed that this error is rather insignificant, especially in two-
dimensional model of ``hot spots'', which is of the main interest for the 
physics of HTSC-cuprates.

The aim of the present paper is to perform an analysis of number of important
aspects of this ``nearly exactly'' solvable model, mainly for two-dimensional
case. We shall consider both the case of short range order fluctuations of 
CDW(SDW,AFM)--type in the ``hot spots '' model \cite{Sch,SchP}, as well as 
the possible applications of this model in the scenario of superconducting
fluctuations\cite{Levin,Tch1,Ren} (SC short range order), in particular for
the most interesting case of $d$-wave pairing. Besides the general discussion
of reliability of the formal scheme of Refs.\cite{C1,C2,C79,C91,S91,Sch,SchP}, 
we shall perform detailed calculations of the spectral densities and 
one-electron density of states both for the ``hot spots'' model\cite{Sch,SchP},
and in the scenario of SC--fluctuations.


\section{``Hot--Spots'' model.}

\subsubsection{Model description and ``nearly exact'' solution for the 
Green's function.}

The model of ``nearly-antiferromagnetic'' Fermi-liquid \cite{Mont1,Mont2}
is based upon the picture of well developed fluctuations of AFM short range
order in a wide region of phase diagram shown in Fig.1. In this model the
effective interaction of electrons with spin fluctuations is described via
dynamic spin susceptibility $\chi_{\bf q}(\omega)$, which is determined 
mainly from the fit to NMR experiments\cite{Mont2}:  
\begin{equation} V_{eff}({\bf q},\omega)=g^2\chi_{\bf q}(\omega)\approx 
\frac{g^2\xi^2}{1+\xi^2({\bf q-Q})^2-i\frac{\omega}{\omega_{sf}}} \label{V} 
\end{equation} 
where $g$ is coupling constant, $\xi$--correlation length of spin 
fluctuations, ${\bf Q}=(\pi/a,\pi/a)$--vector of antiferromagnetic ordering 
in insulating phase, $\omega_{sf}$--characteristic frequency of spin
fluctuations, $a$--lattice spacing.

As dynamic spin susceptibility $\chi_{\bf q}(\omega)$ has peaks at the wave
vectors around $(\pi/a,\pi/a)$ there appear ``two types'' of quasiparticles 
---``hot quasiparticles'' with momenta in the vicinity of ``hot spots'' on 
the Fermi surface (Fig.2) and ``cold'' quasiparticles with momenta on the
other parts of the Fermi surface, e.g. around diagonals of the Brillouin 
zone $|p_x|=|p_y|$ \cite{Sch,SchP}. These terms are connected with the fact
that quasiparticles from the vicinity of ``hot spots'' are strongly
scattered on the vector of the order of ${\bf Q}$ by spin fluctuations
(\ref{V}), while for quasiparticles with momenta far from ``hot spots'' this
interaction is relatively weak.

In the following we shall consider the case of high enough temperatures when
$\pi T \gg \omega_{sf}$ which corresponds to the region of ``weak pseudogap''
in Fig.1 \cite{Sch,SchP}. In this case spin dynamics is irrelevant and we can
limit ourselves to static approximation:
\begin{equation}
V_{eff}({\bf q})=\tilde\Delta^2\frac{\xi^2}{1+\xi^2({\bf q-Q})^2}
\label{Vef}
\end{equation}
where $\tilde\Delta$--effective energy parameter, which in the model of AFM
fluctuations can be written as \cite{SchP}:
\begin{equation}
\tilde\Delta^2=g^2T\sum_{m{\bf q}}\chi_{\bf q}(i\omega_m)=
g^2<{\bf S}_{i}^{2}>/3
\label{dd}
\end{equation}
where ${\bf S}_i$--spin on the lattice site ($Cu$ ion in highly conducting
$CuO_2$--plane for HTSC-cuprates). In the following we shall consider
$\tilde\Delta$ (as well as $\xi$) as some phenomenological parameter,
determining the effective width of the pseudogap.

We can greatly simplify all calculations if instead of (\ref{Vef}) we use
another form of model interaction (cf. analogous model in \cite{Kam}):
\begin{equation}
V_{eff}({\bf q})=\Delta^2\frac{2\xi^{-1}}{\xi^{-2}+(q_x-Q_x)^2}
\frac{2\xi^{-1}}{\xi^{-2}+(q_y-Q_y)^2}
\label{Veff}
\end{equation}
where $\Delta^2=\tilde\Delta^2/4$. In fact (\ref{Veff}) is qualitatively
similar to (\ref{Vef}) and differs from it very slightly in most important
region of $|{\bf q-Q}|<\xi^{-1}$.

Consider the first order in $V_{eff}$ correction to electron self-energy
shown in Fig.3:
\begin{equation}
\Sigma(\varepsilon_n{\bf p})=\sum_{\bf q}V_{eff}({\bf q})
\frac{1}{i\varepsilon_{n}-\xi_{\bf p+q}}
\label{sig}
\end{equation}
Main contribution to the sum over ${\bf q}$ comes from the region close to
${\bf Q}=(\pi/a,\pi/a)$. Then we can write:
\begin{equation}
\xi_{\bf p+q}=\xi_{\bf p+Q+k}\approx \xi_{\bf p+Q}+{\bf v_{\bf p+Q}k}
\label{linsp}
\end{equation}
where $v^{\alpha}_{\bf p+Q}=\frac{\partial\xi_{\bf p+Q}}{\partial
p_{\alpha}}$, and performing integration over ${\bf k}$, we get:
\footnote
{In Refs.\cite{Sch,SchP} another but similar in spirit to (\ref{Veff}) form
of effective interaction was used: $V_{eff}(\bf k)=\Delta^2\frac{2\xi^{-1}}
{\xi^{-2}+k_{\|}^2}\frac{2\xi^{-1}}{\xi^{-2}+k_{\bot}^2}$, where  
$k_{\|(\bot)}$ --projection of $\bf k$ parallel (perpendicular) to 
${\bf v}_ {\bf p+Q}$, so that result analogous to (\ref{sigm}) takes the 
form:  
\begin{eqnarray}
\Sigma(\varepsilon_n{\bf p})=\frac{\Delta^2}{i\varepsilon -
\xi_{\bf p+Q}+i|{\bf v}_{\bf p+Q}|\kappa sign\varepsilon_n}
\nonumber
\end{eqnarray}}
\begin{equation}
\Sigma(\varepsilon_n{\bf p})=\frac{\Delta^2}{i\varepsilon_n-\xi_{\bf p+Q}
+(|v^x_{\bf p+Q}|+|v^y_{\bf p+Q}|)\kappa sign\varepsilon_n}
\label{sigm}
\end{equation}
where $\kappa=\xi^{-1}$.

The spectrum of ``bare'' (free) quasiparticles can be taken in the form
\cite{Sch,SchP}:
\begin{equation}
\xi_{\bf p}=-2t(\cos p_xa+\cos p_ya)-4t^{'}\cos p_xa\cos p_ya
\label{spectr}
\end{equation}
where $t$--nearest neighbor transfer integral,   $t^{'}$--second nearest
neighbor transfer integral on the square lattice, $\mu$ is chemical potential.
In the analysis of real HTSC-systems in Refs.\cite{Sch,SchP} it was assumed
e.g. for $YBa_2Cu_3O_{6+\delta}$ that $t=0.25 eV$,\ $t^{'}=-0.45t$ , 
while $\mu$ was fixed by hole concentration. Below we shall see that it is of
considerable interest to study our problem for different relations between 
$t$ and $t^{'}$.

Consider second order corrections to self-energy shown in Fig.4. Using
(\ref{Veff}) we obtain:
\begin{eqnarray}
\Sigma(a)=\Delta^4\int\frac{d{\bf k_1}}{\pi^2}\int\frac{d\bf k_2}{\pi^2}
\frac{\kappa}{\kappa^2+k_{1x}^2}\frac{\kappa}{\kappa^2+k_{1y}^2}
\frac{\kappa}{\kappa^2+k_{2x}^2}\frac{\kappa}{\kappa^2+k_{2y}^2} \nonumber \\
\frac{1}{i\varepsilon_n-\xi_{\bf p+Q}-v^x_{\bf p+Q}k_{1x}-v^y_{\bf p+Q}k_{1y}}
\frac{1}{i\varepsilon_n-\xi_{\bf p}-v^x_{\bf p}(k_{1x}+k_{2x})
-v^y_{\bf p}(k_{1y}+k_{2y})}
\nonumber \\
\frac{1}{i\varepsilon_n-\xi_{\bf p+Q}-v^x_{\bf p+Q}k_{1x}-v^y_{\bf p+Q}k_{1y}}
\label{sig1}
\end{eqnarray}

\begin{eqnarray}
\Sigma(b)=\Delta^4\int\frac{d{\bf k_1}}{\pi^2}\int\frac{d\bf k_2}{\pi^2}
\frac{\kappa}{\kappa^2+k_{1x}^2}\frac{\kappa}{\kappa^2+k_{1y}^2}
\frac{\kappa}{\kappa^2+k_{2x}^2}\frac{\kappa}{\kappa^2+k_{2y}^2} \nonumber \\
\frac{1}{i\varepsilon_n-\xi_{\bf p+Q}-v^x_{\bf p+Q}k_{1x}-v^y_{\bf p+Q}k_{1y}}
\frac{1}{i\varepsilon_n-\xi_{\bf p}-v^x_{\bf p}(k_{1x}+k_{2x})
-v^y_{\bf p}(k_{1y}+k_{2y})}
\nonumber \\
\frac{1}{i\varepsilon_n-\xi_{\bf p+Q}-v^x_{\bf p+Q}k_{2x}-v^y_{\bf p+Q}k_{2y}}
\label{sig2}
\end{eqnarray}
where we have used the spectrum (\ref{spectr}), from which, in particular,
it follows that $\xi_{\bf p+2Q}=\xi_{\bf p}$, ${\bf v}_{\bf p+2Q}=
{\bf v}_{\bf p}$ for ${\bf Q}=(\pi/a,\pi/a)$. If signs of $v_{\bf p}^x$ and 
$v_{\bf p+Q}^x$, and of $v_{\bf p}^y$ and $v_{\bf p+Q}^y$ are the same, we
can see that integrals in (\ref{sig1}) and (\ref{sig2}) are fully determined
by poles of Lorentzians, which determine the interaction with fluctuations.
After elementary contour integration we obtain:
\footnote{In the model of $V_{eff}$ used in Ref.\cite{Sch,SchP}, 
for the case of ${\bf v}_{\bf p}{\bf v}_{\bf p+Q}>0$, in a similar way we 
get: 
\begin{eqnarray} 
\Sigma(a)=\Sigma(b)=\Delta^4\frac{1}{[i\varepsilon_n-\xi_{\bf p+Q}+
i|{\bf v}_{\bf p+Q}|\kappa]^2}\frac{1}{i\varepsilon_n-\xi_{\bf p+Q}+
i2|{\bf v}_{\bf p}|(|\cos\phi|+|\sin\phi|)\kappa}
\nonumber
\end{eqnarray}
where $\phi$--angle between $\bf v_{\bf p}$ and $\bf v_{\bf p+Q}$.}
\begin{equation}
\Sigma(a)=\Sigma(b)=\frac{1}{[i\varepsilon_n-\xi_{\bf p+Q}+i(|v_{\bf p+Q}^x|
+|v_{\bf p+Q}^y|)\kappa]^2}\frac{1}{i\varepsilon_n-\xi_{\bf p}+
i2(|v_{\bf p}^x|+|v_{\bf p}^y|)\kappa}
\label{sig3}
\end{equation}
Here and in the following we assume $\varepsilon_n>0$. It is easy to
convince ourselves that in case of velocity projections of the same sign
we can similarly calculate contributions of any higher order diagrams.
Accordingly, the contribution of an arbitrary diagram for electron 
self-energy of $N$-th order in interaction (\ref{Veff}) has the form:
\begin{equation}
\Sigma^{(N)}(\varepsilon_n{\bf p})=\Delta^{2N}\prod_{j=1}^{2N-1}
\frac{1}{i\varepsilon_n-\xi_{j}+in_jv_j\kappa}
\label{Ansatz}
\end{equation}
where $\xi_j=\xi_{\bf p+Q}$ and $v_j=|v_{\bf p+Q}^{x}|+|v_{\bf p+Q}^{y}|$ for
odd $j$ and $\xi_j=\xi_{\bf p}$ and $v_{j}=|v_{\bf p}^x|+|v_{\bf p}^{y}|$
for even $j$. Here $n_j$ is the number of interaction lines, enveloping $j$-th
Green's function in a given diagram.

In this case any diagram with intersecting interaction lines is actually
equal to some diagram of the same order with noncrossing interaction lines.
Thus in fact we can consider only diagrams with nonintersecting interaction
lines, taking into account diagrams with intersecting lines introducing
additional combinatorial factors into interaction vertices. This method was
first introduced (for another problem) in a paper by Elyutin \cite{Ely} and 
used for one-dimensional model of the pseudogap state in Refs.
\cite{C79,C91,S91}.

As a result we obtain the following recursion relation for one-electron
Green's function (continuous fraction representation)\cite{C79,C91,S91}):
\begin{equation}
G^{-1}(\varepsilon_n\xi_{\bf p})=G_{0}^{-1}(\varepsilon_n\xi_{\bf p})-
\Sigma_{1}(\varepsilon_n\xi_{\bf p})
\label{G}
\end{equation}
\begin{equation}
\Sigma_{k}(\varepsilon_n\xi_{\bf p})=\Delta^2\frac{v(k)}
{i\varepsilon_n-\xi_k+ikv_k\kappa-\Sigma_{k+1}(\varepsilon_n\xi_{\bf p})}
\label{rec}
\end{equation}
where $\xi_k=\xi_{\bf p+Q}$ and $v_k=|v_{\bf p+Q}^x|+|v_{\bf p+Q}^y|$  for
odd values of $k$,\ $\xi_k=\xi_{\bf p}$ and $v_k=|v_{\bf p}^x|+
|v_{\bf p}^y|$ for even $k$. Combinatorial factor:
\begin{equation}
v(k)=k
\label{vcomm}
\end{equation}
corresponds to our case of commensurate fluctuations with
${\bf Q}=(\pi/a,\pi/a)$ \cite{C79}. It is not difficult to analyze also the
case of incommensurate fluctuations when $\bf Q$ is not ``locked'' to the
period of inverse lattice. In this case diagrams with interaction lines
enveloping  odd number of vertices are significantly smaller than
diagrams with interaction lines enveloping even number of vertices.
Thus we must consider only these last diagrams \cite{C1,C2,C79,C91,S91}.
Recurrence relation has the same form (\ref{rec}), but diagram combinatorics
and factors $v(k)$ change\cite{C79}:
\begin{equation}
v(k)=\left\{\begin{array}{cc}
\frac{k+1}{2} & \mbox{for odd $k$} \\
\frac{k}{2} & \mbox{for even $k$}
\end{array} \right.
\label{vincomm}
\end{equation}
In Refs.\cite{Sch,SchP} a spin-structure of effective interaction within
the model of ``nearly antiferromagnetic'' Fermi-liquid was taken into 
account (spin-fermion model of Ref.\cite{SchP}). This leads to more
complicated combinatorics of diagrams in the commensurate case with 
${\bf Q}=(\pi/a,\pi/a)$. Spin-conserving part of the interaction
gives formally commensurate combinatorics, while spin-flip scattering is
described by diagrams with combinatorics of incommensurate type (``charged'' 
random field in terms of Ref.\cite{SchP}). As a result the recurrence 
relation for the Green's function is again of the form of(\ref{rec}), but the
combinatorial factor $v(k)$ is now\cite{Sch,SchP}:  
\begin{equation} 
v(k)=\left\{\begin{array}{cc}
\frac{k+2}{3} & \mbox{for odd $k$} \\
\frac{k}{3} & \mbox{for even $k$}
\end{array} \right.
\label{vspin}
\end{equation}

As we noted above, the solution of the form of (\ref{rec}) can be obtained
only in the case of coinciding signs of velocity projections
$v_{\bf p+Q}^x(v_{\bf p+Q}^y)$ and $v_{\bf p}^x(v_{\bf p}^y)$. Below we shall
analyze situations when this is really so. In case of different signs of 
these projections integrals of the type of (\ref{sig1}) and (\ref{sig2}), 
corresponding to higher order corrections, can not be calculated in such a
simple way as above, because contributions from the poles of the Green's
functions become relevant. In this case, instead of simple answers like
(\ref{sig3}) rather cumbersome expressions appear and, even more importantly,
disappears the fundamental (for our method) fact of equality of wide range
of diagrams with crossing and noncrossing interaction lines, which actually
allows us to classify higher order contributions and obtain an ``exact''
solution (\ref{rec}). This problem is important only for the case of finite
correlation lengths of fluctuations $\xi=\kappa^{-1}$, while in the limit of
$\xi\to\infty$($\kappa\to 0$) the exact solution for the Green's function
is independent of velocities ${\bf v}_{\bf p}$ and ${\bf v}_{\bf p+Q}$ and
can be easily obtained in analytic form by methods of Refs.\cite{C1,C2} 
(cf. also\cite{SchP}). In one-dimensional model considered in Refs.
\cite{C1,C2,C79,C91,S91} the signs of corresponding velocity projections are
always different (these correspond to ``left'' and ``right'' moving 
electrons). This problem was stressed in a recent paper \cite{Tch2}. In
Appendix A we analyze these difficulties in detail for one-dimensional case
and show that the ``Ansatz'' of the type of (\ref{Ansatz}) used in Refs.
\cite{C79,C91,S91} for the contributions of higher order diagrams as well as
solution of the form of (\ref{rec}) in fact give us very good approximation 
to an exact solution even in the case of velocity projections of different
signs. Obviously this solution is exact in the limits of
$\xi\to\infty$($\kappa\to 0$) and $\xi\to 0$($\kappa\to\infty$), and
guarantees rather good (qualitatively) description in the region of finite
correlation lengths.

\subsubsection{Analysis of the ``bare'' energy-spectrum.}
For the ``bare'' energy spectrum (\ref{spectr}) we can easily find conditions
(relations between $t$,$t'$ and $\mu$), when solution (\ref{rec}) is exact.
First of all, let us define the region of parameters $t$,$t'$ and $\mu$,
when ``hot spots'' on the Fermi surface exist, i.e. the condition of
existence of points connected with vector ${\bf Q}=(\pi/a,\pi/a)$. 
If ${\bf p}=(p_x,p_y)$ is the position of a ``hot spot'' on the Fermi
surface, then ${\bf p+Q}=(p_x+\pi/a,p_y+\pi/a)$ must also belong to the Fermi
surface, so that for the spectrum (\ref{spectr}) we must have:
\begin{eqnarray}
-2t(\cos p_xa+\cos p_ya)-4t'\cos p_xa\cos p_ya -\mu=0 \nonumber \\
2t(\cos p_xa+\cos p_ya)-4t'\cos p_xa\cos p_ya -\mu=0
\label{hsp}
\end{eqnarray}
Then the condition of existence of ``hot spots'' becomes:
\begin{equation}
\cos p_ya=-\cos p_xa \qquad\mbox{and}\qquad \cos^2 p_xa=\frac{\mu}{4t'}
\label{hotsp}
\end{equation}
Thus, the ``hot spots'' on the Fermi surface exist if:
\begin{equation}
0\leq\frac{\mu}{4t'}\leq 1
\label{hotspot}
\end{equation}
Define now the region of parameters $t$,$t'$ and $\mu$ where (\ref{rec}) is
exact requiring the positivity of products $v_{\bf p}^x v_{\bf p+Q}^x$ 
and $v_{\bf p}^y v_{\bf p+Q}^y$. We have:  
\begin{eqnarray} 
v_{\bf p}^x=\frac{\partial\xi_{\bf p}}{\partial p_x}=
2t\sin p_xa+4t'\sin p_xa\cos p_ya \nonumber\\
v_{\bf p}^y=\frac{\partial\xi_{\bf p}}{\partial p_y}=
2t\sin p_ya+4t'\sin p_ya\cos p_xa \nonumber\\
v_{\bf p}^xv_{\bf p+Q}^x=16t'^2\sin^2 p_xa[\cos^2 p_ya-(\frac{t}{2t'})^2]
\nonumber\\
v_{\bf p}^yv_{\bf p+Q}^y=16t'^2\sin^2 p_ya[\cos^2 p_xa-(\frac{t}{2t'})^2]
\label{vv}
\end{eqnarray}
It is easily seen that for the existence of points on the Fermi surface,
where the projections of velocities are of the same signs, it is necessary
to fulfil $|t'/t|>1/2$. We are mainly interested in the region around the
``hot spots'', where with an account of(\ref{hotsp}) we have:
\begin{equation}
v_{\bf p}^xv_{\bf p+Q}^x=v_{\bf p}^yv_{\bf p+Q}^y=4t^2(1-\frac{\mu}{4t'})
(\frac{\mu t'}{t^2}-1)
\label{vvvv}
\end{equation}
Thus the projections of velocities in ``hot spots'' have the same sign if:
\begin{equation}
\frac{\mu t'}{t^2}>1
\label{vsign}
\end{equation}
The same condition obviously guarantees the validity of inequality
${\bf v}_{\bf p}{\bf v}_{\bf p+Q}>0$ which is necessary for exactness of
(\ref{rec}) in the model of Refs.\cite{Sch,SchP}.

In Fig.5 we show (dashed region) the region of parameters where ``hot spots''
exist $(0\leq\mu/4t'\leq 1)$ as well as the region where velocity 
projections in their vicinity are of the same sign $(\mu t'>1)$. In Fig.6 we
show the Fermi surfaces, defined by the ``bare'' spectrum (\ref{spectr}),
for different values of chemical potential $\mu$ (band fillings) when these
conditions are either satisfied or not. 
 
\subsubsection{Spectral density and density of states.}

Consider one-electron spectral density:
\begin{equation}
A(E{\bf p})=-\frac{1}{\pi}Im G^R(E{\bf p})
\label{spd}
\end{equation}
where $G^R(E{\bf p})$ is retarded Green's function, obtained by the usual
analytic continuation of (\ref{G}) to the real axis of energy $E$.
In Fig.7 we show energy dependencies of $A(E{\bf p})$ obtained from
(\ref{G}),(\ref{rec}) for different variants of combinatorial factors
(\ref{vcomm}),(\ref{vincomm}),(\ref{vspin}). For $t'/t=-0.6$  and 
$\mu/t=-1.8<t/t'=1.666$ projections of velocities in ``hot spots'' are of the
same sign and relation (\ref{rec}) defines Green's function exactly.
We can see that in incommensurate case (\ref{vincomm}) (Fig.7(a)) as well as for
combinatorics of spin-fermion model (\ref{vspin}) (Fig.7(c)) the spectral 
density at the ``hot spot'' demonstrates clearly non Fermi-liquid behavior 
(for large enough values of correlation length $\xi$). Note also that for both 
cases the numerical values of spectral density are very close. In the case of
commensurate combinatorics (\ref{vcomm}) (Fig.7(b)) precisely at the 
``hot spot'' the spectral density has a single peak and, in this sense, 
is similar to that of the Fermi-liquid even for large values of $\xi$. 
However, even in the nearest neighborhood of ``hot spot'' this spectral density 
acquires two peak structure (``shadow'' band) of non Fermi-liquid type for 
large enough values of $\xi$ (see insert in Fig.7(b)).

Far from ``hot spots'' velocity projections are, in general, of different
signs even if condition (\ref{vsign}) is satisfied. Accordingly, the
recurrence relation (\ref{rec}) for the Green's function is non exact here.
\footnote{
At the same time, with the growth of $\xi$ more narrow vicinity of the 
``hot spot'' becomes important and our approximation is more and more
accurate.}
However, from discussion in Appendix A it becomes clear that our ``Ansatz''
(\ref{Ansatz}) and solution (\ref{rec}) in fact somehow overestimate
the role of finite correlation length $\xi$. There we also provide slightly
different variant of solution (\ref{altkap}), which somehow underestimates
this role. 
Inserts in Fig.7 we show energy dependencies of spectral density far
from the ``hot spot'' for different combinatorics of diagrams (\ref{vcomm}),
(\ref{vincomm}), (\ref{vspin}). 

For comparison we also show the spectral
density (for incommensurate case) obtained with an ``Ansatz of alternating
$\kappa$''(\ref{altkap}) which happens to be very close to that obtained
from (\ref{G}), (\ref{rec}). This confirms rather high accuracy of (\ref{rec})
for arbitrary momenta close to the Fermi surface.

In Fig.8 we show energy dependencies of spectral density for different
combinatorics (\ref{vcomm}),(\ref{vincomm}),(\ref{vspin}) at the ``hot spot''
in the case of $t'/t=-0.4$, which, according to Ref.\cite{Sch,SchP},
corresponds to $YBa_2Cu_3O_{6+\delta}$. For this value of $t/t'$ even at the
``hot spot'' velocity projections are of different signs. However,
spectral density (for incommensurate case) obtained from the solution with
``alternating'' $\kappa$ (dashed line in Fig.8(a)) is seen to be very close
to that obtained from (\ref{Ansatz}). This shows that ``Ansatz'' (\ref{Ansatz})
and solution (\ref{rec}) produce results which are quantitatively close to an
exact solution. Let us stress once
again that our solution (\ref{rec}) is exact both for $\xi\to\infty$ and
$\xi\to 0$, while in the region of finite $\xi$ it provides rather good
interpolation.

Consider now one-electron density of states:
\begin{equation}
N(E)=\sum_{\bf p}A(E,{\bf p})=-\frac{1}{\pi}\sum_{\bf p}ImG^R(E{\bf p})
\label{dos}
\end{equation}
which is determined by the integral of spectral density $A(E{\bf p})$ over 
the Brillouin zone. We have seen above that though for some topologies of
the ``bare'' Fermi surface (band fillings) we can guarantee the same signs
of velocity projections close to the ``hot spots'', this is not so in 
general case far from ``hot spots'', so that solution (\ref{rec}), based
upon our ``Ansatz'' (\ref{Ansatz}), is only approximate.
Accordingly, the use of (\ref{rec}) to calculate the density of states
with (\ref{dos}) leads also to a kind of approximation. In Fig.9 we show
densities of states obtained from (\ref{G}),(\ref{rec}),(\ref{dos}), with the
use of the ``bare'' spectrum (\ref{spectr}) for different diagram 
combinatorics (\ref{vcomm}),(\ref{vincomm}),(\ref{vspin}), for the values of 
$t'/t=-0.4$ (Fig.9(a)) and $t'/t=-0.6$ (Fig.9(b)). We can see that for
$t'/t=-0.4$ there appears some dip in the density of states (pseudogap).
This dip is relatively weakly dependent on the value of correlation length
$\xi$ (see insert in Fig.9(b)). 
If the band filling is appropriate and the Fermi level $\mu$
is somewhere in  this energy region there are also ``hot spots'' at the
Fermi surface. For $t'/t=-0.6$ the region of existence of ``hot spots'' is
rather wide, but the pseudogap in the density of states is practically
unobservable. We can see only the obvious smearing of Van-Hove singularity
which is present in the absence of fluctuations.


\section{Model of ``superconducting'' fluctuations.}

\subsubsection{Model description and solution for the Green's function.}

Pseudogap phenomena can also be probably explained using the ideas of
fluctuation Cooper pairing at temperatures higher than superconducting
transition temperature $T_c$\cite{R,Gesh,EK,Levin}. Let us consider the
simplest possible model approach to this problem. In Fig.10(a) we show the
self-energy diagram of the first order in fluctuation propagator of Cooper
pairs for $T>T_c$. Anticipating the possibility of both usual $s$-wave
and $d$-wave pairing, characteristic of HTSC-systems, we introduce the
pairing interaction of the simplest (separable) form:
\begin{equation}
V({\bf p,p'})=-Ve(\phi)e(\phi')
\label{Vsc}
\end{equation}
where $\phi$ is polar angle determining the direction of electronic momentum
${\bf p}$ in the plane, while for $e(\phi)$ we assume model dependence
\cite{Bork,Fehr}:
\begin{equation}
e(\phi)=\left\{\begin{array}{cc}
1 & \mbox{$s$-wave pairing}\\
\sqrt{2}\cos(2\phi) & \mbox{$d$-wave pairing}
\end{array} \right.
\label{e}
\end{equation}
Interaction constant $V$ is as usual assumed to be non zero in some energy
layer around the Fermi surface. Then the self-energy corresponding to
Fig.10(a) takes the form:
\begin{equation}
\Sigma(\varepsilon_n{\bf p})=\sum_{m{\bf p}}V_{eff}(i\omega_m{\bf q})
G(i\omega_m-i\varepsilon_n,{\bf -p+q})
\label{scsig}
\end{equation}
where effective interaction with SC--fluctuations can be written as:
\begin{equation}
V_{eff}(i\omega_m{\bf q})=-\frac{Ve^2(\phi)}
{1-VT\sum_{n{\bf p}}G_{0}(i\varepsilon_n{\bf p})G_{0}(i\omega_m-
i\varepsilon_n,{\bf -p+q})e^2(\phi)}
\label{Vesc}
\end{equation}
In the following we assume SC--fluctuations static
\footnote{Static approximation here is valid for
$\pi T\gg\omega_{sc}=8(T-T_c)/\pi$, which is formally analogous to the
condition of $\pi T\gg\omega_{sf}$ used in the ``hot spot'' model above.
Here it is well satisfied if the system is close enough to superconducting
transition}so that in (\ref{scsig1}) we can limit ourselves only to the term
with $\omega_m=0$. Then effective interaction can be written as:
\begin{equation}
V_{eff}({\bf q})\approx -\frac{\tilde\Delta^2 e^2(\phi)}{\xi^{-2}(T)+
{\bf q}^2}
\label{Vefsc}
\end{equation}
where
\begin{equation}
\xi(T)=\frac{\xi_0}{\sqrt{\frac{T-T_c}{T_c}}} \qquad \mbox{;} \qquad
\xi_{0}\approx0.18\frac{v_F}{T_c}
\label{xi}
\end{equation}
is the usual coherence length of superconductor, $\tilde\Delta^2=\frac{1}
{N(E_F)\xi_0^2}$ ($N(E_F)$--density of states at the Fermi level $E_F$).
Surely, within the simplest BCS-like model used here, we have 
$\tilde\Delta\approx2\pi^2T_c(T_c/E_F)\sim\Delta_0(\Delta_0/E_F)\ll\Delta_0$
(where $\Delta_0$ is superconducting energy gap at $T=0$) and an obvious
problem to explain the experimentally observable scale of pseudogap anomalies
appears. However, in the following we again consider both $\xi$ and
$\tilde\Delta$ as some phenomenological parameters to be determined from
experiment on HTSC-systems and not from naive BCS-like model. 

Analogously to transition from (\ref{Vef}) to (\ref{Veff}) we introduce
instead of (\ref{Vefsc}) the model interaction of the form:  
\begin{equation} 
V_{eff}({\bf q})=-\Delta^2 e^2(\phi)\frac{2\xi^{-1}}{\xi^{-2}+q_x^2} 
\frac{2\xi^{-1}}{\xi^{-2}+q_y^2}
\label{Veffsc}
\end{equation}
where $\Delta^2=\tilde\Delta^2/4$. Quantitatively this is close enough to
(\ref{Vefsc}) and leads to great simplification of calculation allowing us
to classify contributions of higher order diagrams. In this case the first
order contribution of diagram in Fig.10(a) is:
\begin{equation}
\Sigma^{(1)}(\varepsilon_n{\bf p})=\frac{\Delta^2 e^2(\phi)}
{i\varepsilon_n+\xi_{\bf p}+i(|v_x|+|v_y|)\kappa sign\varepsilon_n}
\label{scsig1}
\end{equation}
where $v_x=v_f\cos\phi$,$v_y=v_F\sin\phi$,$\kappa=\xi^{-1}$. Second order
contribution from Fig.10(b) is:
\begin{eqnarray}
\Sigma^{(2)}(\varepsilon_n{\bf p})=(\Delta^2e^2(\phi))^2
\int\frac{d q_{1x}}{\pi}\frac{\kappa}{\kappa^2+q_{1x}^2}
\int\frac{d q_{1y}}{\pi}\frac{\kappa}{\kappa^2+q_{1y}^2} 
\int\frac{d q_{1x}}{\pi}\frac{\kappa}{\kappa^2+q_{2x}^2}
\int\frac{d q_{1y}}{\pi}\frac{\kappa}{\kappa^2+q_{2y}^2} \nonumber\\
\frac{1}{(i\varepsilon_n+\xi_{\bf p}-{\bf v_1q_1})^2} 
\frac{1}{i\varepsilon_n-\xi_{\bf p}-{\bf v_2q_1}-{\bf v_2q_2}}
\label{scsig2}
\end{eqnarray}
where ${\bf v}_1=-{\bf v}_2={\bf v}_F$. In fact we can easily see that in 
this problem we have practically the same rules of diagram technique as in
the ``hot spot'' model, but with combinatorics of incommensurate case.
This last fact is obvious from the topology of interaction line (fluctuation
propagator of Cooper pairs) in diagram of Fig.10(a) --- it is seen that in
higher orders only those diagrams exist in which interaction lines envelop
only even number of interaction vertices. The expression of (\ref{scsig2}) 
is quite analogous to that of (\ref{sig1}), but the signs of velocity
projections in denominators of Green' functions here are always different:
${\bf v}_1=-{\bf v}_2$. Thus in diagrams of higher orders there appear
contributions not only from Lorentzians of interaction, but also from the
Green's functions. However (in view of discussion in Appendix A) we can
estimate contributions of higher order diagrams using the ``Ansatz'' of the
type of (\ref{Ansatz}), i.e. calculate all integrals e.g. in (\ref{scsig2}), 
as if velocity projections ${\bf v}_1$ and ${\bf v}_2$ are of the same sign,
but in {\em answer} just put ${\bf v}_1=-{\bf v}_2={\bf v}_F$.  
Then we again obtain the recurrence relation for the Green's function of
the type of (\ref{rec}):  
\begin{equation} \Sigma_{k}(\varepsilon_n\xi_{\bf 
p})=\frac{\Delta^2e^2(\phi)v(k)} {i\varepsilon_n-(-1)^k\xi_{\bf 
p}+ikv_F\kappa(|\cos\phi|+|\sin\phi|)- \Sigma_{k+1}(\varepsilon_n\xi_{\bf p})} 
\label{recsc} 
\end{equation} 
where $v(k)$ is defined in (\ref{vincomm}). 
Surely, this relation (\ref{recsc}) is not exact, but again it gives 
exact results for the limits of $\kappa\to 0$($\xi\to\infty$) and 
$\kappa\to\infty$($\xi\to 0$) and provides rather good (quantitatively)
interpolation between these limits for the case of finite correlation lengths.

\subsubsection{Spectral density and density of states.}

In Fig.11(a) we show energy dependencies of the spectral density 
$A(E{\bf p})$ for one-particle Green's function (\ref{spd}), calculated from
(\ref{recsc}) for different values of polar angle $\phi$, determining the
direction of electronic momentum in the plane (we take here $|{\bf p}|=p_F$),
for the case of fluctuations of $d$-wave pairing. It is clearly seen that in
the vicinity of the point $(\pi/a,0)$ in Brillouin zone this spectral density
is non Fermi-liquid like (pseudogap). As vector ${\bf p}$ rotates in the
direction of the zone diagonal the two peak structure gradually disappears
and spectral density transforms to the typical Fermi-liquid like with a
single peak, which is narrows as $\phi$ approaches $\pi/4$. Analogous
transformation of the spectral density takes place as correlation length
$\xi$ becomes smaller. 

In Fig.11(b) we also show the evolution of the product $f(E)A(E{\bf p})$ 
(where $f(E)$ is fermi distribution) which is essentially the parameter
measured in ARPES experiments\cite{RC}. Note that curves in Fig.11(b) are
quite similar to those obtained in Refs.\cite{Sch,SchP} within the ``hot
spots'' model. This picture of Fermi-surface destruction following from this
calculations is qualitatively shown in Fig.12 and is very similar to
experimental data obtained e.g. in Ref.\cite{Nor} for 
$Bi_2Sr_2CaCu_2O_{8+\delta}$.

In case of fluctuation pairing of $s$-wave type the pseudogap appears
isotropically on the whole Fermi-surface and spectral density is non
Fermi-liquid everywhere  for the case of large enough correlation lengths
$\xi$ of SC--fluctuations.

In Fig.13 we present the results of calculations of one-electron density of
states with the help of (\ref{recsc}) both for the case of $s$-wave pairing
(Fig.13(a)) and $d$-wave pairing (Fig.13(b)), for different values of
correlation length of SC--fluctuations. We see that in the case of $d$-wave
pairing the pseudogap is the density of states is naturally not so deep as
in $s$-wave case, even for large enough correlation lengths. At the same time
it is seen that the pseudogap in the density of states in the model of
SC--fluctuations is nevertheless much more expressive than in the model of
``hot spots'' due to AFM--fluctuations.


\section{Conclusion.}

In this paper we have studied ``nearly exactly'' solvable models of the 
pseudogap state of the electronic spectrum of two-dimensional systems,
based upon alternative scenarios of its origin --- the picture of 
fluctuations of ``dielectric'' (AFM, SDW, CDW) type, leading to the model of
of ``hot spots'', and the picture of fluctuational Cooper pairing above
$T_C$ (SC--fluctuations). The term ``nearly exactly'' solvable  means that
in this approach we can sum all Feynman diagram series for one-particle
Green's function (and in fact also for two-particle Green' function
\cite{C91,S91}), using for the higher order diagrams an {\em approximate}
``Ansatz'' (\ref{Ansatz}).  As shown in Appendix A and also on numerical
examples in the main part of the paper this ``Ansatz'' guarantees rather
good (quantitatively) approximation to an exact solution in the region of
finite correlation lengths of fluctuations of short range order  $\xi$, 
while in the limits of $\xi\to\infty$  and $\xi\to 0$ our solution is exact.

Calculation of the spectral densities shows that within both scenarios
we can obtain rather attractive (in the sense of possible comparison with the
experimental data on cuprates) picture of ``destruction'' of Fermi-liquid
behavior on specific (``hot'') parts of the Fermi-surface, with persistence
of Fermi-liquid state on the rest (``cold'') part of the Fermi-surface.
This non Fermi-liquid behavior is due to the strong scattering of electrons
on fluctuations of short range order and, in general, is more visible with
the growth of correlation length $\xi$. At the same time there are definite
differences between these two scenarios which can be used, in principle,
in the analysis of real experimental situation. In particular, in the
``hot spots'' model (AFM--fluctuations) the pseudogap in the density of
states is relatively weak (cf. Fig.9), while in the model of 
SC--fluctuations the pseudogap in the density of states is more visible
(cf. Fig.13). The model of ``dielectric'' AFM--fluctuations is more
attractive even from simplest consideration of the phase diagram of Fig.1 ---
pseudogap anomalies are mainly observed in the underdoped region and are
apparently more intensive for systems which are closer to dielectric
(AFM) state. It is obvious that precisely in this region we can expect more
important role of fluctuations of ``dielectric'' (AFM) type and the growth of
corresponding correlation length $\xi$. It is rather difficult to
imagine why in this region of the phase diagram SC--fluctuations may become
more important, this apparently must take place somewhere close to the
optimal (corresponding to highest $T_c$) doping. Also in SC--scenario we
have an obvious problem of characteristic scales (on temperature and energy) 
of pseudogap anomalies, which can not be solved within simple BCS-like 
theory, and requires some new microscopic approaches\cite{Gesh,Levin}.  
Our models are useful for the analysis within both scenarios of pseudogap
formation irrespective of microscopic picture, because they are based on
rather general (semi phenomenological) form of correlation function of
fluctuations of short range order.

Authors are grateful to O.V.Tchernyshyov for preliminary information on
his analysis of one-dimensional model.

This work is partly supported by Russian Basic Research Foundation under the
grant No.96-02-16065, as well as by projects No.IX.1 of the State Program 
``Statistical Physics'' and No.96-051 of the State Program on HTSC of the
Russian Ministry of Science. 

\newpage

\appendix

\section{Analysis of one-dimensional model.}

Let us consider in more detail the use of the ``Ansatz'' (\ref{Ansatz}) to
estimate the contributions of higher order diagrams. We shall limit 
ourselves to the analysis of one-dimensional model \cite{C79,C91,S91},
because in one-dimension the problem is most serious\cite{Tch2}. We are
interested in the vicinity of Fermi ``points'' $+p_F$ ¨ $-p_F$, while the
Gaussian fluctuations of short range order scatter electrons by the momentum
of the order of $Q\sim ^+_-2p_F$ from one end of the Fermi ``line'' 
to the opposite with scattering momentum values fixed with precision of the
order of $\xi^{-1}=\kappa$ \cite{C1,C2,C79,C91,S91}. We shall consider the
linearized electronic spectrum: $\xi_{p^-_+p_F}=^+_-v_Fp$ and to shorten
notations put $v_F=1$. Thus our system consists of ``two types'' of
electrons --- those moving to the ``left'' and to the ``right''. It is
convenient to make our analysis in the coordinate representation \cite{Tch2}, 
when the equation of motion of electrons in our model takes the form
\cite{McK,Tch2}:  
\begin{equation} \Bigl(i\hat{1}\frac{\partial}{\partial 
t}-i\hat\sigma_3\frac{\partial} {\partial x}\Bigr)\hat\Psi(t,x)= 
\left(\begin{array}{cc}
0 & \Delta(x)  \\
\Delta^{\star}(x) & 0 
\end{array} \right)\hat\Psi(t,x)
\label{eqmot}
\end{equation}
We limit ourselves with incommensurate fluctuations only when
$\Delta^{\star}(x)\neq\Delta(x)$. Spinor 
$\hat\Psi=\left(\begin{array}{cc}
\psi_{+} & \\
\psi_{-} &
\end{array}\right)$ 
describes ``right'' and ``left'' electrons. Fluctuations $\Delta(x)$ are
supposed to be Gaussian with $<\Delta(x)=0>$ and 
$<\Delta^{\star}(x)\Delta(x')>=|\Delta|^2 exp(-\kappa|x-x'|)$. Free 
propagator in frequency and coordinate representation has the form:
\begin{equation} 
G_0(\varepsilon x)=i\theta(\varepsilon\sigma_3 x)sign(\varepsilon)
exp(i\varepsilon\sigma_3 x)
\label{G0}
\end{equation}
where $\sigma_3=+1$ for the ``right'' particles, $\sigma_3=-1$ for the 
``left''. The particle passing the path of the length $l$ produces a phase
factor $e^{i\varepsilon l}$. During calculation of contribution of a given
diagram it is convenient to change integration variables from coordinates of
interaction vertices $x_k$ to path lengths $l_k$, passed by the particle
between scattering events\cite{Tch2}. It is important to take into account
the fact that these path lengths are not independent as for the given
diagram the total particle displacement $x-x'$ is fixed. The rules of the
diagram technique to calculate $G(\varepsilon,x-x')$ are as follows\cite{Tch2}:
\begin{enumerate}
\item{Electron line of length $l_k$ gives the factor 
$-ie^{il_k(\varepsilon-(-1)^kp)}$.} 
\item{Wavy (interaction) line connecting vertices $m$ and $n$ gives the
factor: 

$|\Delta|^2exp{(-\kappa|x_{m}-x_{n}|)}=|\Delta|^2exp{(-\kappa|\sum_{k=m}
^{n-1}(-1)^kl_k|)}$.}

\item{Over all $l_k$ we must integrate from $0$ to $\infty$.}
\item{Integrate over $p$ with a factor of $e^{ip(x-x')}/2\pi$.} 
\end{enumerate} 
To calculate $G(\varepsilon p)$ just drop the last rule. From these rules
we can see that the finite values of correlation length $\xi= \kappa^{-1}$ 
lead to some damping of given transition amplitude with the displacement
of the particle. The exact accounting of this effect is rather complicated but
we can find some upper and lower bound estimates.
Considering first the obvious inequality:
\begin{equation} exp\Bigl(-\kappa|\sum_{k=m}^{n-1}(-1)^kl_k|\Bigr)> 
exp\Bigl(-\kappa\sum_{k=m}^{n-1}l_k\Bigr)
\label{ineq1}
\end{equation}
and using for the interaction line the r.h.s. expression in (\ref{ineq1}),
we overestimate transition amplitude damping (i.e. effectively overestimate
$\kappa$) for the given diagram. It is easy to see that the use of this 
approximation to calculate a given diagram for the Green's function
leads (in momentum representation) to an extra $\i\kappa$ appearing in every
denominator of Green's function enveloped by an extra interaction line.
This leads to an expression for the given higher order contribution of the
form of (\ref{Ansatz}) (cf.\cite{Tch2}). For example, diagram shown in Fig.14
yields (we assume $\varepsilon>0$, $\delta=0^{+}$):  
\begin{equation} 
\Delta G(\varepsilon p)=\Delta^4\frac{1}{\varepsilon-p+i\delta}
\Bigl(\frac{1}{\varepsilon+p+i\kappa}
\frac{1}{\varepsilon-p+2i\kappa}
\frac{1}{\varepsilon+p+i\kappa}\Bigr)
\frac{1}{\varepsilon-p+i\delta}
\label{dG1}
\end{equation}
which is analogous to (\ref{sig1}),(\ref{sig3}). On the other hand we can
take the inequality:
\begin{equation}
exp\Bigl(-\kappa|\sum_{k=m}^{n-1}(-1)^kl_k|\Bigr)<
exp\Bigl(-\kappa\sum_{k=m}^{n-1}(-1)^{k-m}l_k\Bigr)
\label{ineq2}
\end{equation}
and use for the interaction line the r.h.s. expression in (\ref{ineq2}).
This will lead to some underestimation of damping effect in the given
transition amplitude (i.e. effectively underestimate $\kappa$).
\footnote{It may seem that this choice for the interaction line contribution
can even lead to the growth of transition amplitude in comparison with the
case of $\kappa=0$ and to appearance of some divergences, but this is not 
so.  As we consider only incommensurate case here, where the interaction line
envelopes  only even number of interaction vertices (i.e. an odd number of
$l_k$), the choice of the sign in the exponent after removing the modulous is
determined by dominance of even or odd $l_k$. This leads to effective 
damping of any diagram in higher orders.} 
In particular, for diagram in Fig.14 the contribution of interaction lines
is:
\begin{equation}
e^{-\kappa l_2}e^{-\kappa|l_1-l_2-l_3|}\rightarrow
e^{-\kappa l_2}e^{-\kappa(l_1-l_2+l_3)}=e^{-\kappa(l_1+l_3)}
\label{inl}
\end{equation}
In momentum representation this yields:
\begin{equation}
\Delta G(\varepsilon p)=\Delta^4\frac{1}{\varepsilon-p+i\delta}
\Bigl(\frac{1}{\varepsilon+p+i\kappa}
\frac{1}{\varepsilon-p+i\delta}
\frac{1}{\varepsilon+p+i\kappa}\Bigr)
\frac{1}{\varepsilon-p+i\delta}
\label{dG2}
\end{equation}
Analysis of higher order diagrams shows that in this case contributions of
all diagrams of the $N$-th order are equal and in the momentum representation
we have (``Ansatz of alternating $\kappa$''):
\begin{equation}
G_N(\varepsilon p)=|\Delta|^{2N}\frac{1}{(\varepsilon-p+i\delta)^{N+1}}
\frac{1}{(\varepsilon+p+i\kappa)^N}
\label{Gn}
\end{equation}
Then the whole series is easily summed analogously to the case of
$\kappa=0$ \cite{C1,C2} and we obtain the Green's function in the form:
\begin{equation}
G^R(\varepsilon p)=\sum_{N=0}^{\infty}N!G_N(\varepsilon p)=
\int_{0}^{\infty}d\zeta e^{-\zeta}\frac{\varepsilon+p+i\kappa}
{(\varepsilon-p+i\delta)(\varepsilon+p+i\kappa)-\zeta|\Delta|^2}
\label{Galt}
\end{equation}
From this expression we can easily calculate the spectral density or the
density of states:
\begin{equation}
\frac{N(\varepsilon)}{N(E_F)}=\frac{v_F\kappa}{\pi}\int_{-\infty}^{\infty}
d\xi_p\int_{0}^{\infty}d\zeta e^{-\zeta}\frac{\zeta|\Delta|^2}
{(\varepsilon^2-\xi_p^2-\zeta|\Delta|^2)^2+(v_F\kappa)^2(\varepsilon-
\xi_p)^2}
\label{dosalt}
\end{equation}
where we have restored $v_F$ explicitly. In Fig.15 we compare densities of
states for different values of $\kappa$ (correlation length), calculated
using the ``Ansatz of alternating $\kappa$'' and recurrence relations 
similar to (\ref{Ansatz}) in one-dimensional model \cite{C79,C91,S91}.
We see that the results are quantitatively close practically for all values
of $\kappa$. As we noted above our main ``Ansatz''(\ref{Ansatz}),(\ref{dG1}) 
somehow overestimates the role of finite $\kappa$, while the ``Ansatz of
alternating $\kappa$'' (\ref{dG2}) underestimates it. Then it is guessed that
the exact value of the density of states is in fact quite close to those
obtained using these two types of approximation for the higher order 
diagrams. Analogous results can be obtained also for spectral densities.
In fact this means that the results for the main physical quantities
determined by one-electron Green's functions are not strongly dependent on 
the way the finite $\kappa$ enter the expressions for diagrams of higher
order, but the main thing is to take account (at least approximately) of
{\em all} diagrams of perturbation theory with different combinatorics.
In principle this is not very surprising, as the main effect of pseudogap
formation is due essentially to ``backward'' $Q\sim 2p_F$--scattering, which is
accounted for exactly in the limit of $\xi\to\infty$, while the effect of finite
$\kappa$ reduces to rather weak modulation of this random field, leading to the 
damping of its correlator and smearing of the pseudogap. 

Naturally, the ``Ansatz of alternating $\kappa$'' can be written in the form
of recurrence relation of the type of (\ref{rec}) also for two-dimensional,
models discussed in the main part of this paper. For example in ``hot spots''
model we have:
\begin{equation}
\Sigma_{k}(\varepsilon_n\xi_{\bf p})=\Delta^2\frac{v(k)}
{i\varepsilon_n-\xi_k+i\alpha_k v_k\kappa-\Sigma_{k+1}(\varepsilon_n
\xi_{\bf p})} 
\label{altkap} 
\end{equation}
where $\alpha_k=1$ for odd $k$ and $\alpha_k=0$ for even $k$. Other notations
are given above in the main part. Data for the density of states obtained
with the help of (\ref{altkap}) are shown above in Fig.9 and confirm the
our conclusions. Expression similar to (\ref{altkap}) can be easily written
also for the model of SC--fluctuations.

Let us stress that the ``Ansatz of alternating $\kappa$'' is rather formal
and is used only to show that this more or less arbitrary ``approximation'',
underestimating the role of finite $\kappa$ in diagrams of higher orders,
leads to results which are quantitatively close those obtained with the
``Ansatz'' (\ref{Ansatz}),(\ref{dG1}), which overestimates this role.
This last approximation used in Refs.\cite{C79,C91,S91,McK} and in the main
part of this article has much deeper sense. As we have already stated above
this approximation is exact in the vicinity of ``hot spots'' for the values
of parameters of the ``bare'' spectrum $t$,$t'$ and $\mu$  
(topologies of the Fermi surface) which guarantee equal signs of velocity
projections in ``hot spots'' connected by vector ${\bf Q}$.
Analogously in one-dimensional model it is possible to obtain the higher order
contributions in the form similar to (\ref{Ansatz}) or (\ref{dG1}) if we
consider the model of correlator of fluctuations of short range order with the
maximum at some arbitrary scattering vector $Q$ which is much smaller than 
$p_F$, so that (for large enough correlation lengths $\xi$) electrons are
scattered by fluctuations remaining always on one (``left'' or ``right'')
branch of the spectrum. In this case expressions of the type of (\ref{dG1}) 
are exact. After that in final {\em answers} for diagrams of higher orders
we can perform {\em continuation} to $Q\sim 2p_F$ of interest to us, as
the only dependence on $Q$ enters only via the ``bare'' electronic spectrum.
Similarly we can achieve the same result changing appropriately the chemical
potential $\mu$ (band filling).

\newpage

\setlength{\textwidth}{15.0cm}
\begin{center}
{\bf Figure Captions.}
\end{center}
\vskip 0.5cm

Fig.1. Schematic phase diagram of HTSC-cuprates\cite{SchP}. 
For temperatures $T<T^{cr}$ there are well developed fluctuations of AFM
short range order. For $T_{*}<T<T^{cr}$ these fluctuations can be considered
as static.

\vskip 0.3cm

Fig.2. Model of the Fermi surface for HTSC-cuprates. Electronic states around
the intersection points of the Fermi surface with magnetic Brillouin zone
(shown by dashed lines) are strongly interacting with fluctuations of AFM
short range order (``hot spots'').

\vskip 0.3cm

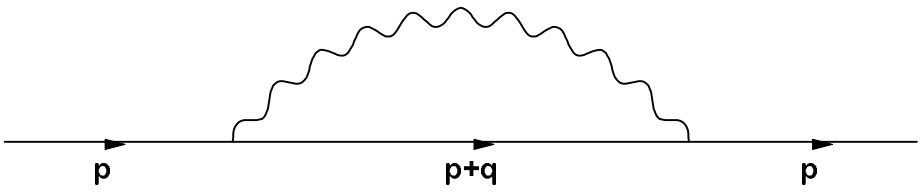
Fig.3. First order self-energy diagram for the electron interacting with
fluctuations of short-range order.

\vskip 0.3cm

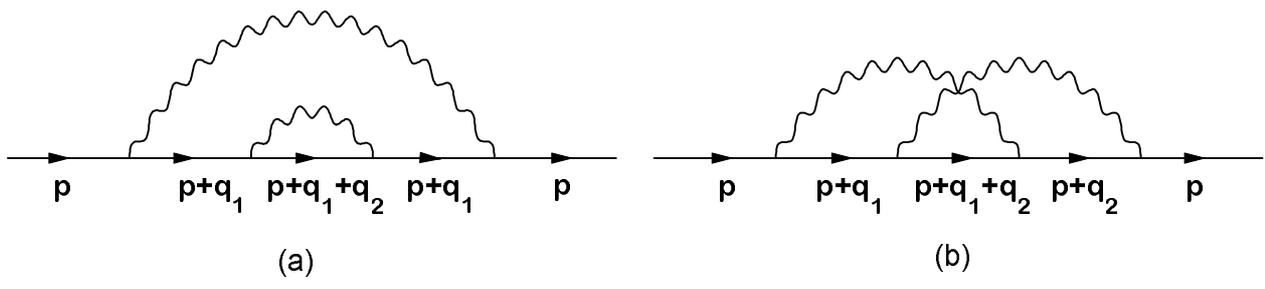
Fig.4. Second order sel-energy diagrams for the electron interacting with
fluctuations of short range order.

\vskip 0.3cm

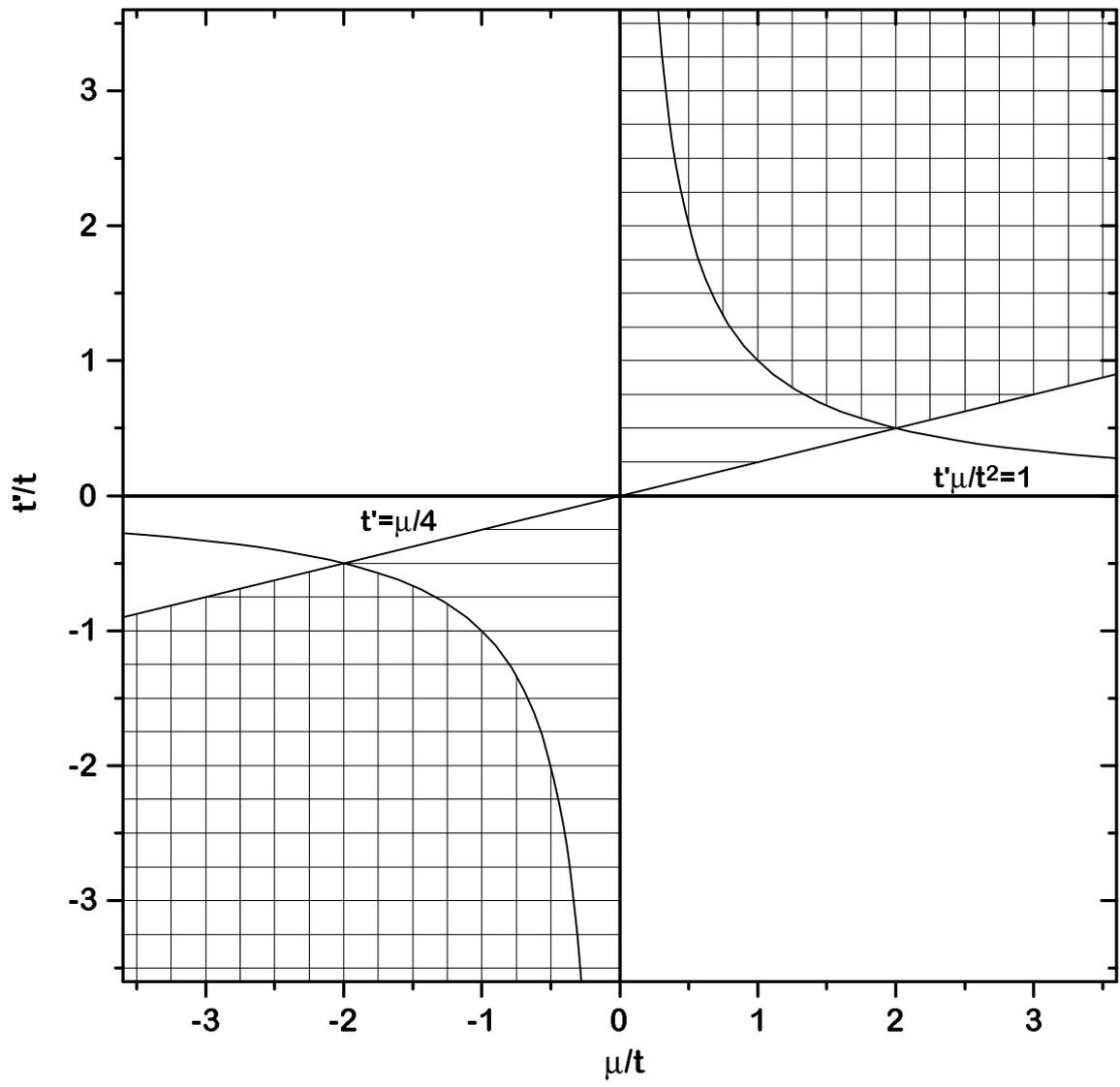
Fig.5. Regions in spectrum parameters space where both ``hot spots'' exist 
(dashed region) and velocity projections in ``hot spots '' are of the same 
sign (doubly dashed region).

\vskip 0.3cm

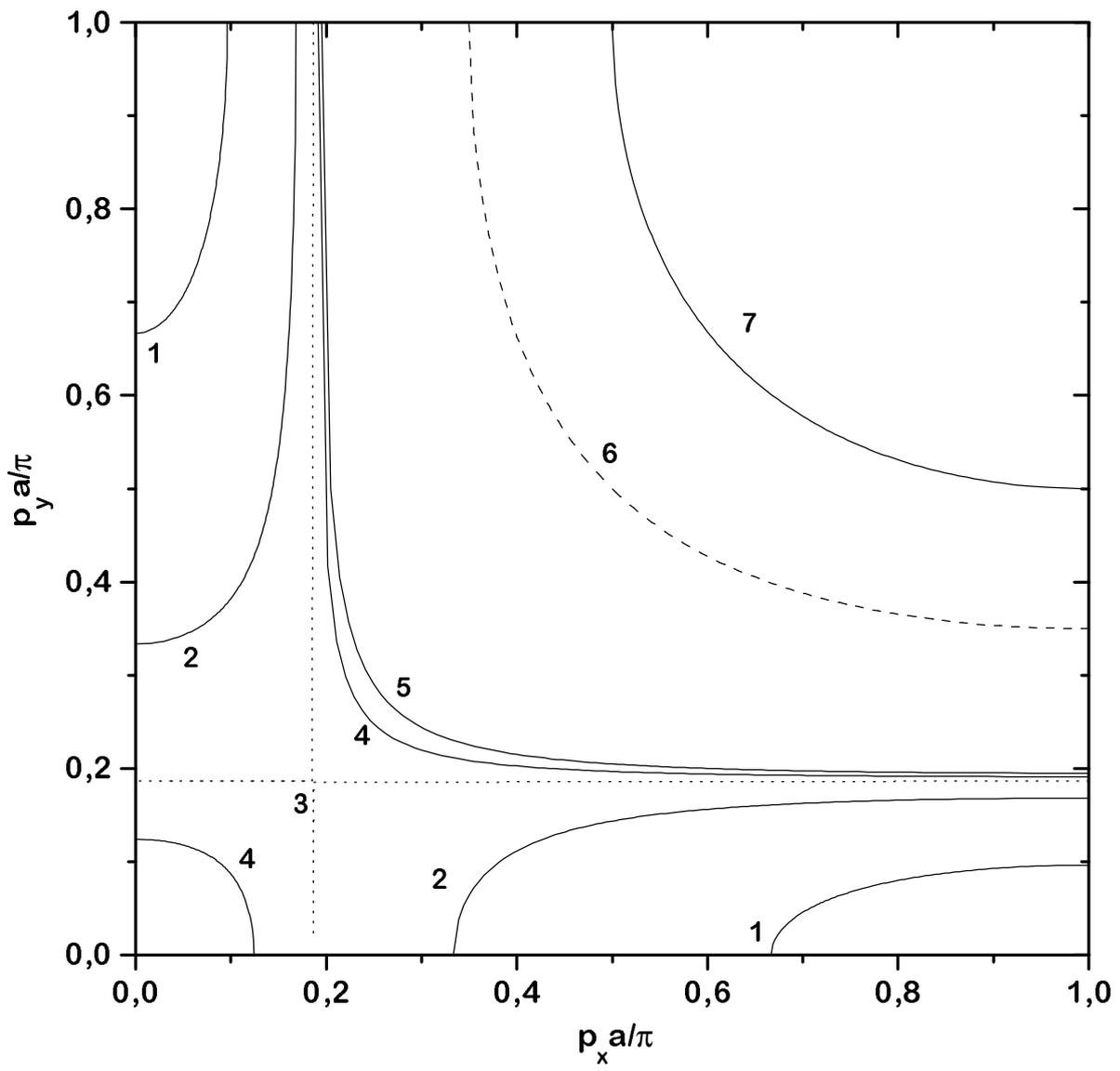
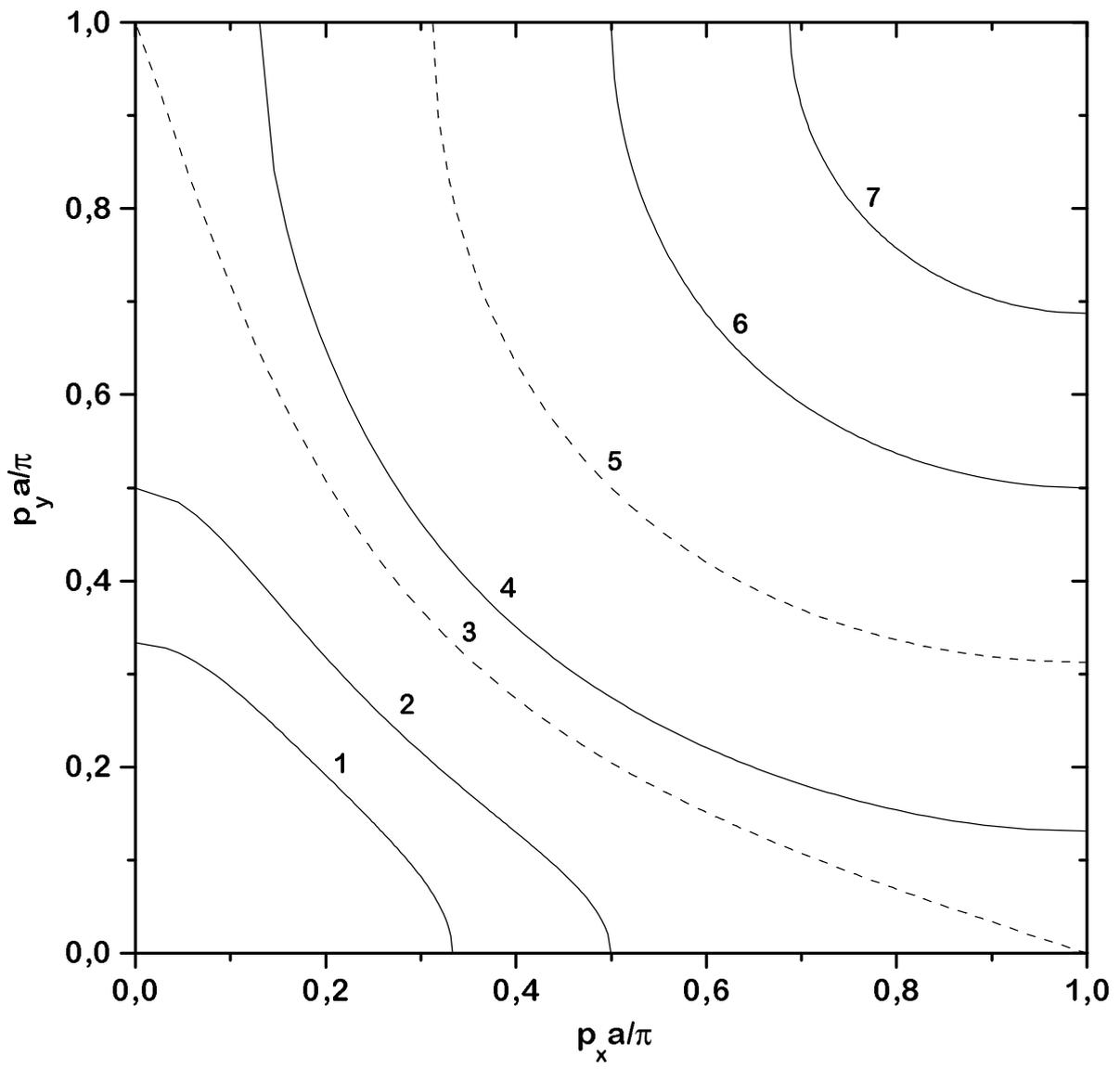
Fig.6. Fermi surfaces defined by the spectrum (\ref{spectr}), for different
values of the chemical potential $\mu$ (band-filling) and parameter $t'/t$.

(a)---case of $t'/t=-0.6$, when (\ref{rec}) is exact close to ``hot spots'':\\ 
$\mu /t=$: (1)--- -2.2; (2)--- -1.8; (3)--- -1.666\ldots;
(4)--- -1.63; (5)--- -1.6; (6)--- 0; (7)--- 2,\\
solution (\ref{rec}) is exact in the vicinity of ``hot spots''(velocity 
projections are of the same sign) for $\mu /t < -1.666\ldots$, ``hot spots'' 
exist for $\mu /t<0$.

(b)---case of $t'/t=-0.4$, characteristic of HTSC-cuprates, when (\ref{rec}) 
is approximate:\\
$\mu /t=$: (1)--- -2.2; (2)--- -2; (3)--- -1.6;
(4)--- -1.3; (5)--- 0; (6)--- 2; (7)--- 4,\\
``hot spots'' exist for $-1.6<\mu /t<0$.

\vskip 0.3cm

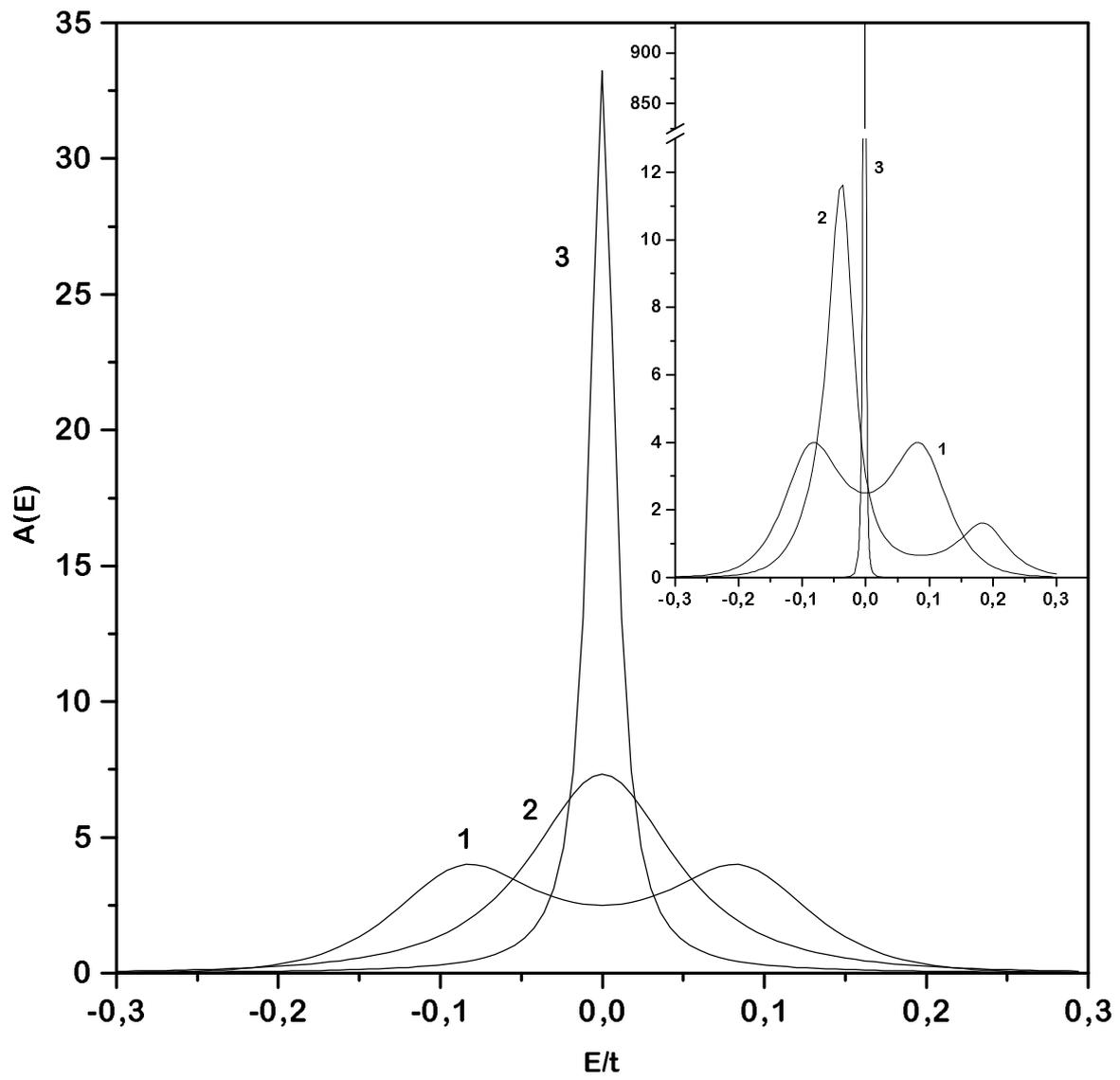
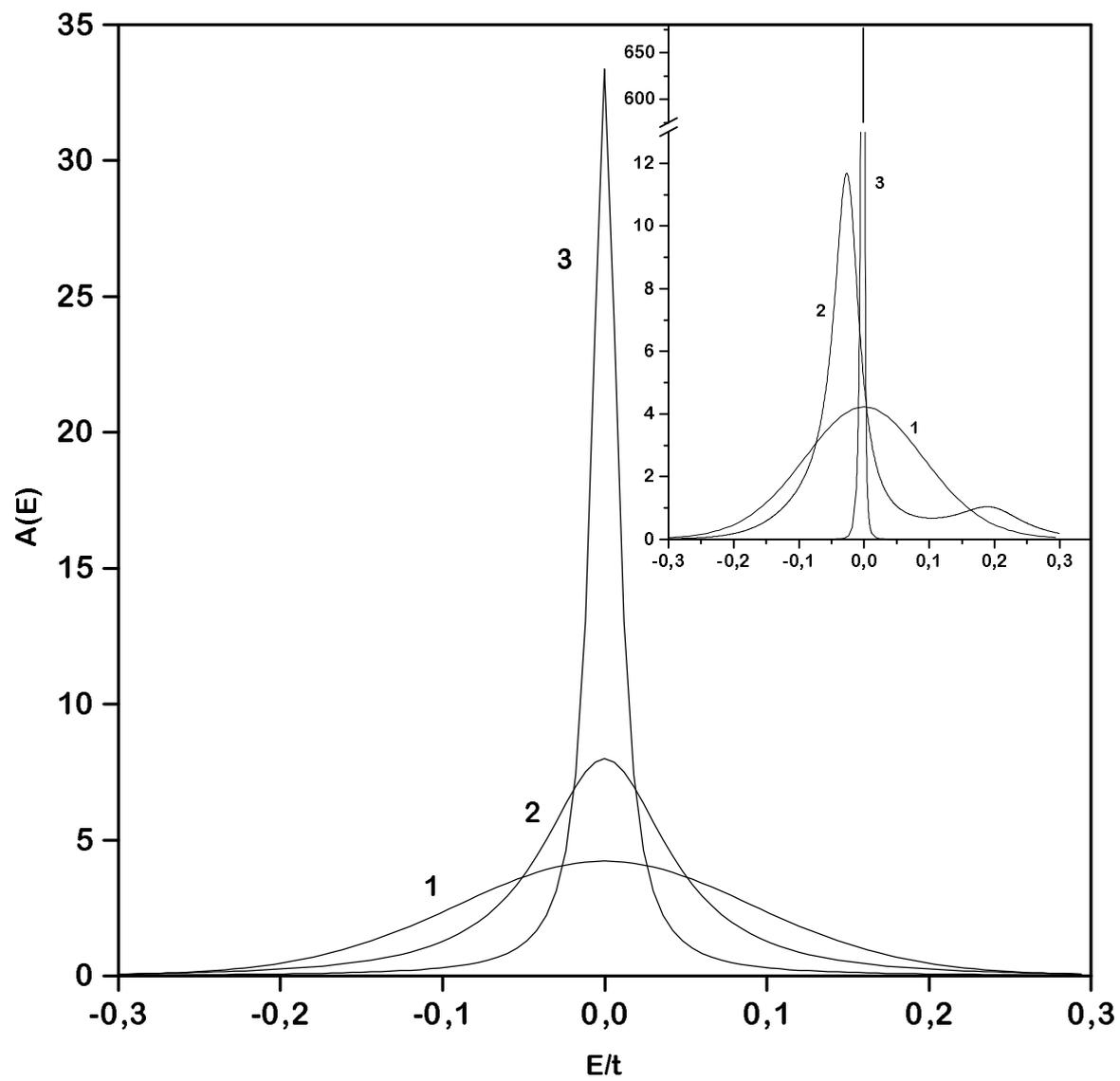
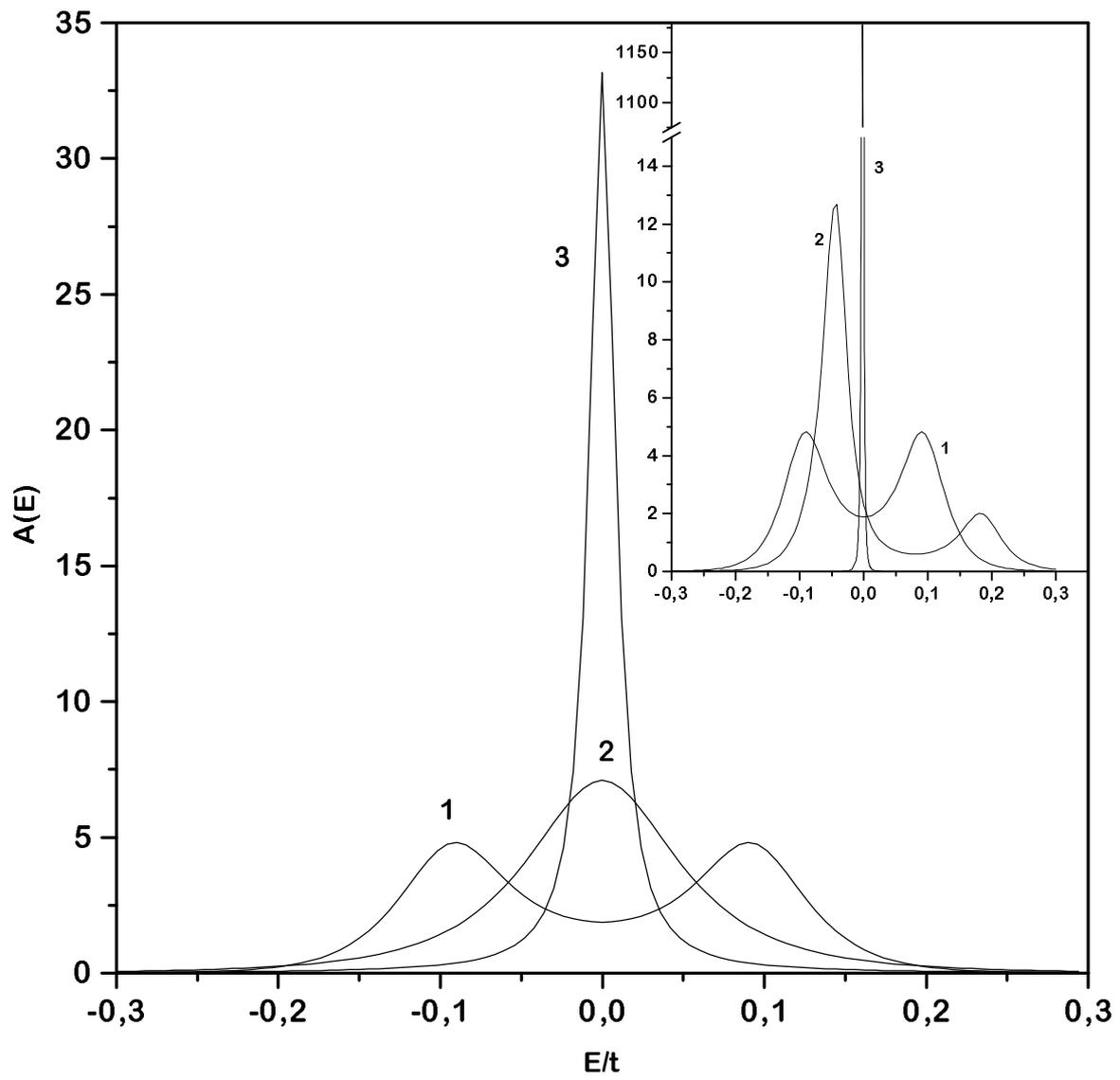
Fig.7. Energy dependencies of the spectral density in the ``hot spot'' 
$(p_xa/\pi=0.1666,p_ya/\pi=0.8333)$ for different diagram combinatorics for the
case of $t'/t=-0.6$, when (\ref{rec}) is exact:

(a)---incommensurate case.

(b)---commensurate case. 

(c)---combinatorics of spin-fermion model.  

Correlation length corresponds to the values of $\kappa$:
(1)---0.01;\ (2)---0.1;\ (3)---0.5,\\
$\Delta=0.1t$.

At the inserts---energy dependencies of spectral density for different
diagram combinatorics for $\kappa a=0.01$

(1)---at the ``hot spot'' $p_xa/\pi =0.1666$, $p_ya/\pi= 0.8333$.

(2)---close to the ``hot spot''  
$p_xa/\pi =0.1663$, $p_ya/\pi =0.8155$.

(3)---far from the ``hot spot''  
$p_xa/\pi =0.0$, $p_ya/\pi =0.333$.

\vskip 0.3cm

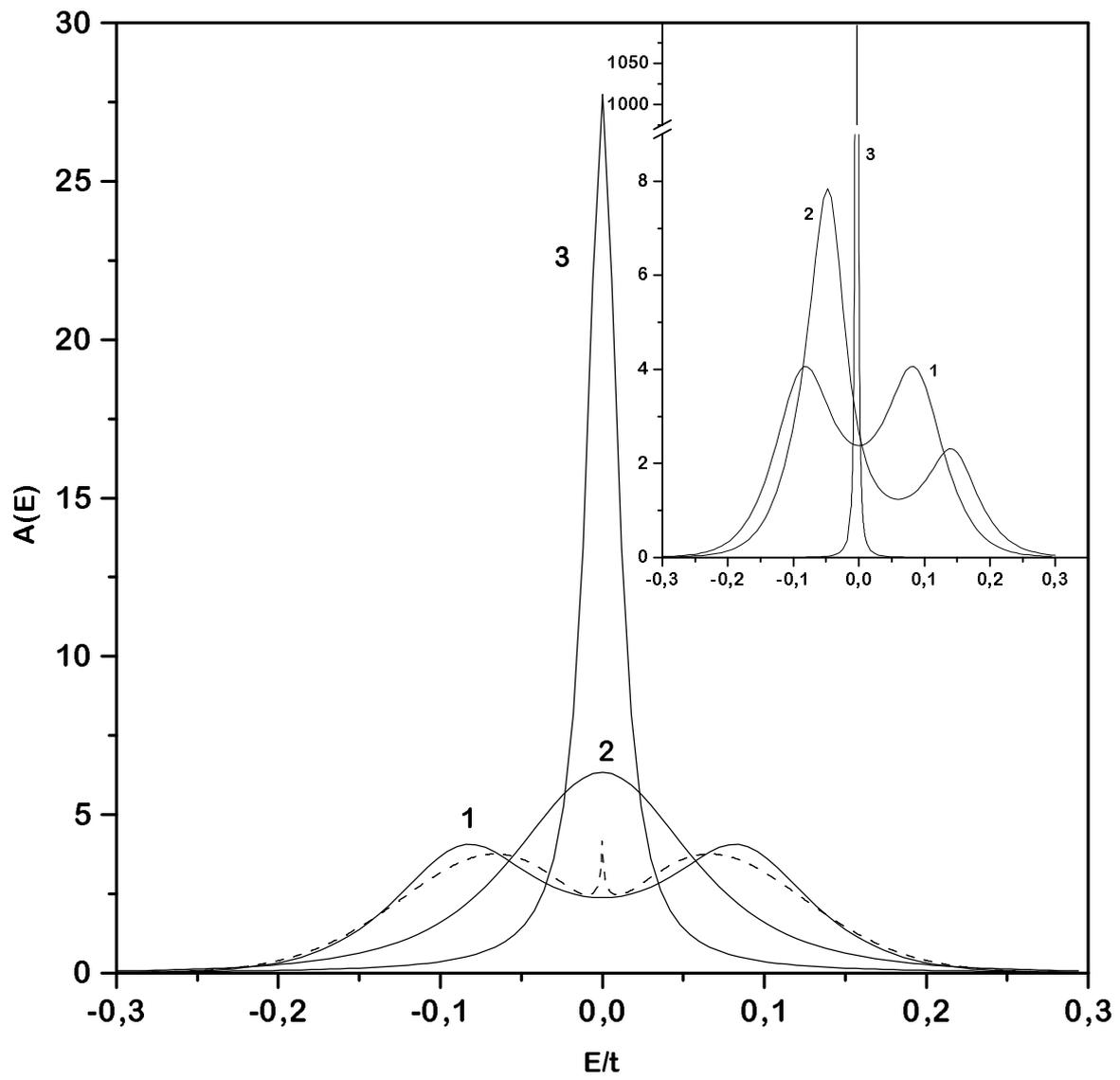
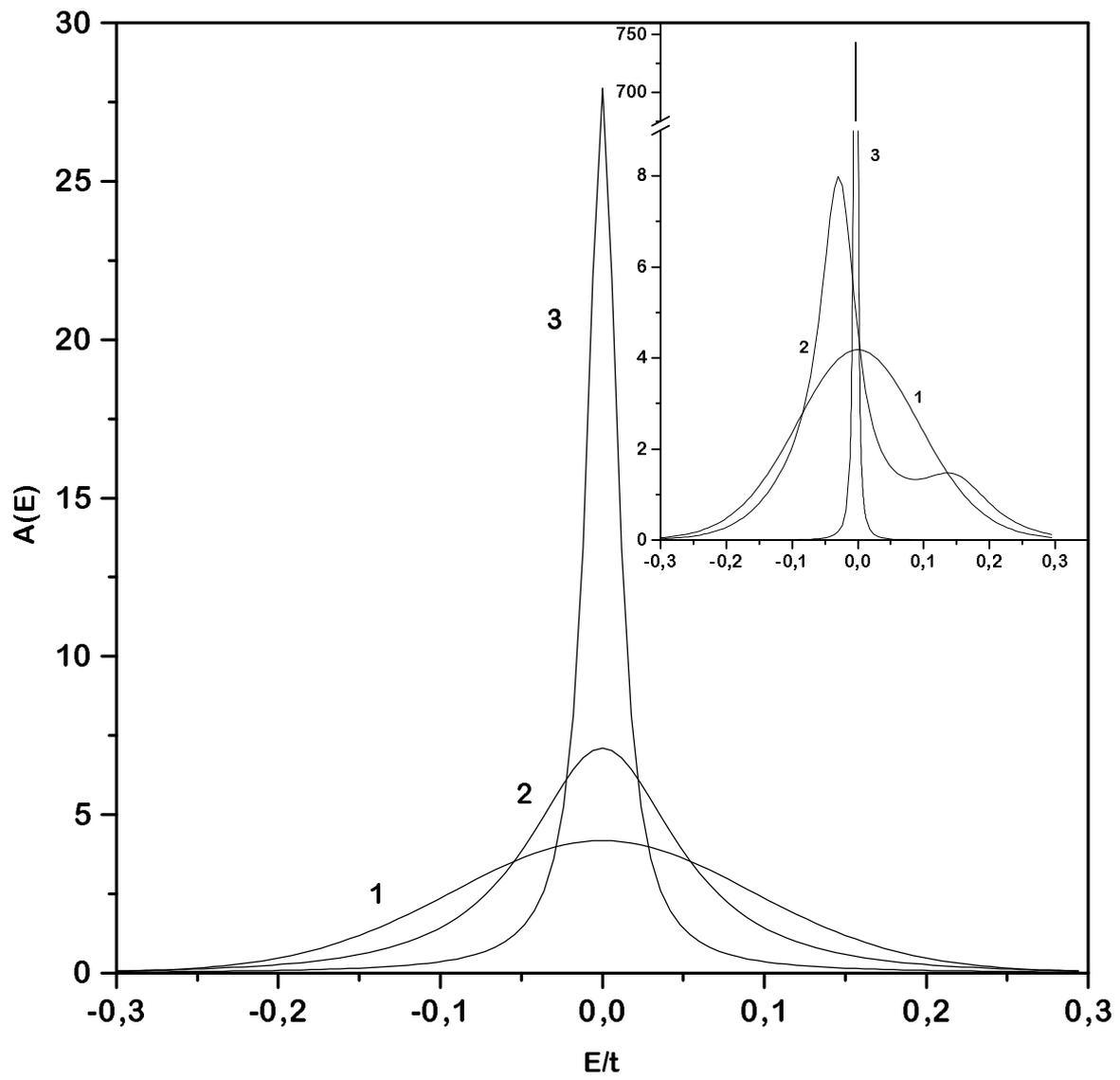
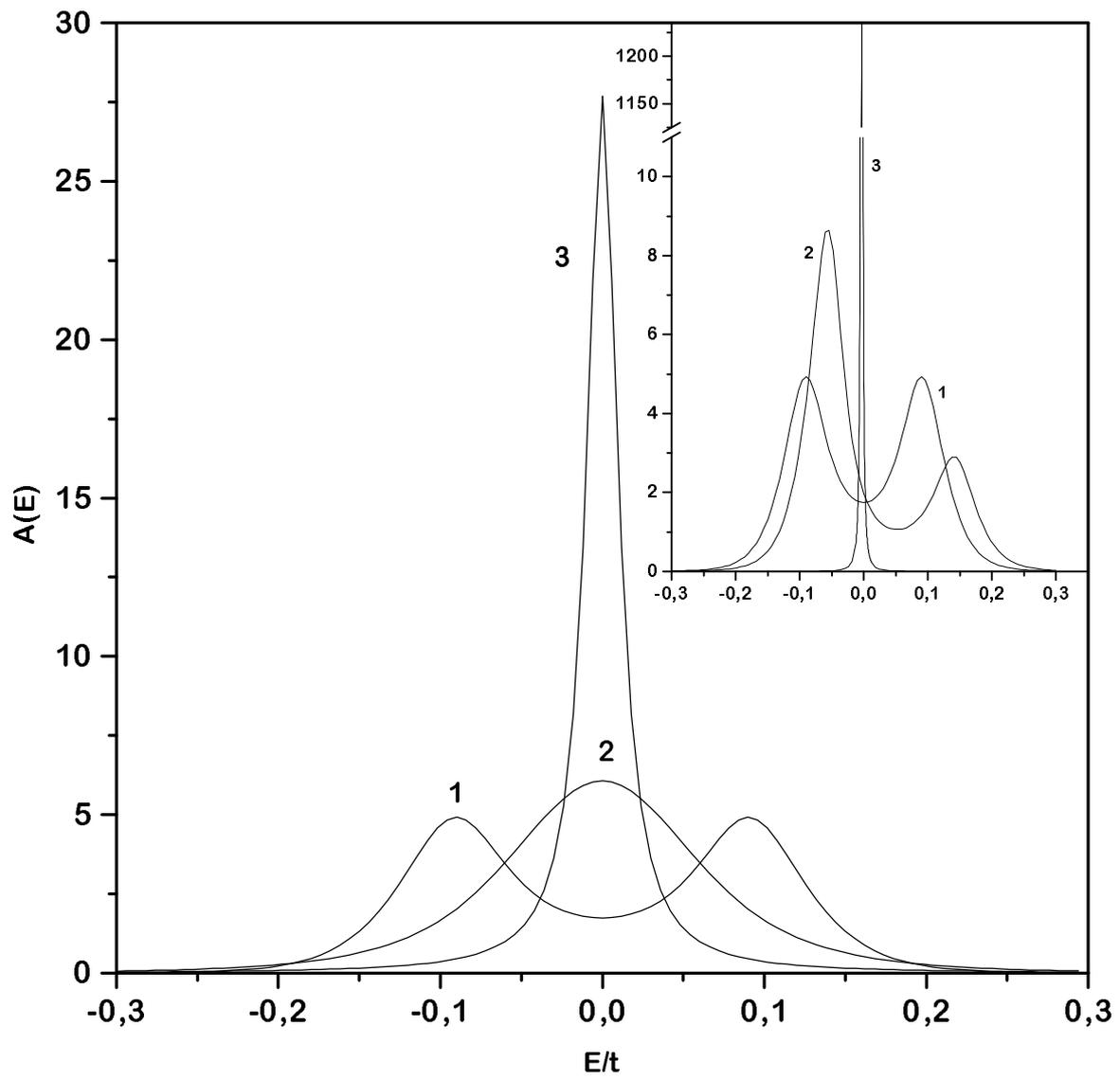
Fig.8. Energy dependencies of the spectral density far from the ``hot spot''
($p_xa/\pi =0.142$, $p_ya/\pi=0.857$)
for  $t'/t=-0.4$, $\mu /t=-1.3$, which is approximately valid 
for HTSC--cuprates:

(a)---incommensurate case.
Dashed line---spectral density for incommensurate case obtained from 
(\ref{altkap}).

(b)---commensurate case. 

(c)---combinatorics of spin-fermion model.  

Correlation length corresponds to the values of $\kappa a$:  
(1)---0.01;\ (2)---0.1;\ (3)---0.5,\\ $\Delta=0.1t$.

At the inserts --- energy dependencies of spectral density for different
diagram combinatorics for $\kappa a=0.01$

(1)---at the ``hot spot'' $p_xa/\pi =0.142$, $p_ya/\pi= 0.857$.

(2)---close to the ``hot spot''  
$p_xa/\pi =0.145$, $p_ya/\pi =0.843$.

(3)---far from the ``hot spot''  
$p_xa/\pi =p_ya/\pi =0.375$.

\vskip 0.3cm

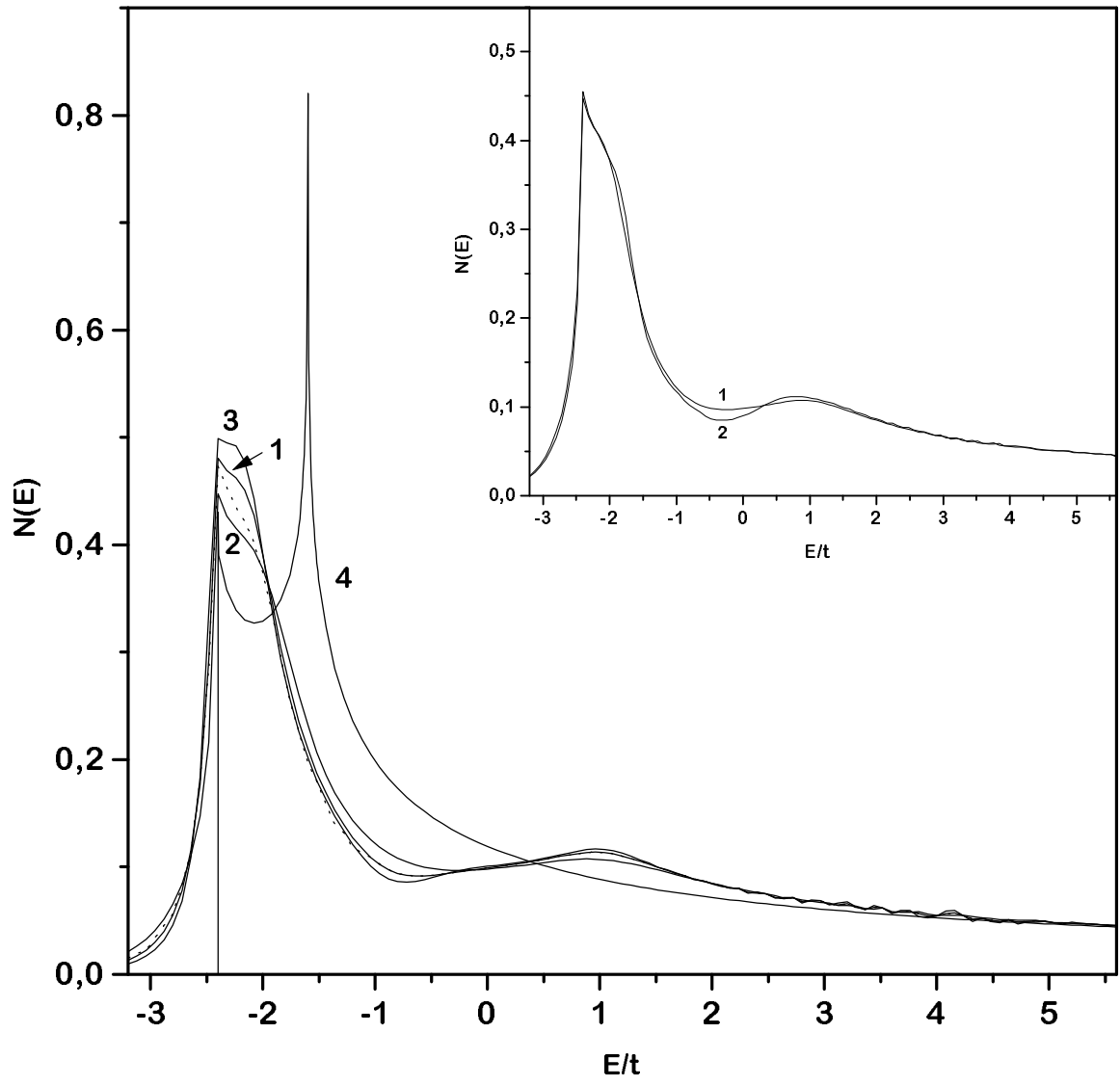
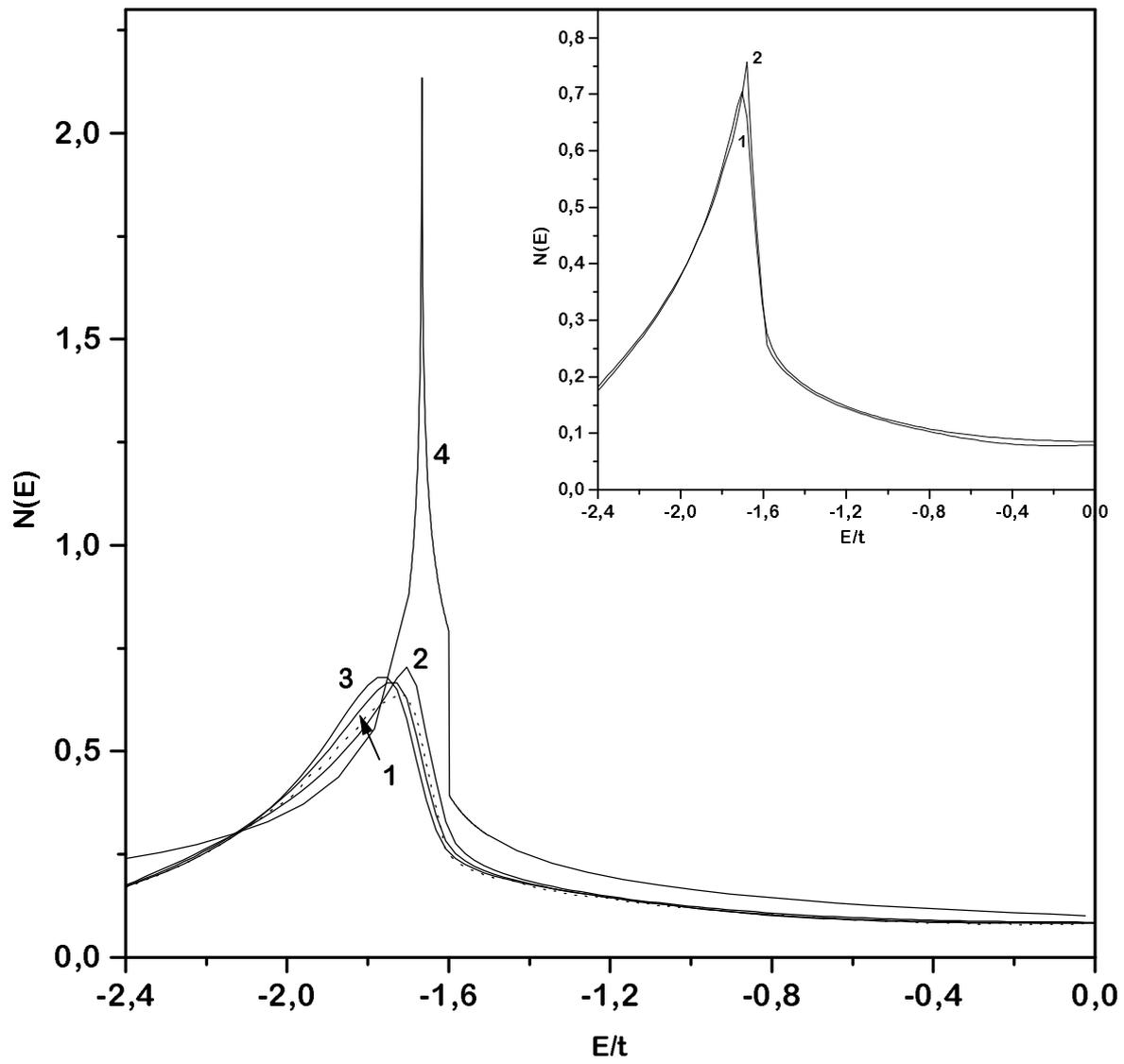
Fig.9. One-electron density of states for different diagram combinatorics
((a)--case of $t'/t=-0.4$,\ $\mu/t=-1.3$; 
(b)--case of $t'/t=-0.6$,$\mu/t=-1.8$):

(1)---incommensurate case.

(2)---commensurate case.

(3)---combinatorics of spin-fermion model.

(4)---in the absence of AFM fluctuations.

Dashed line---incommensurate case, obtained from (\ref{altkap}). 

$\Delta /t=1$, correlation length corresponds to $\kappa a=0.1$.

At the inserts---one-electron density of states for commensurate 
combinatorics  for:

(1)---$\kappa a=0.1$; (2)---$\kappa a=0.01$

\vskip 0.3cm

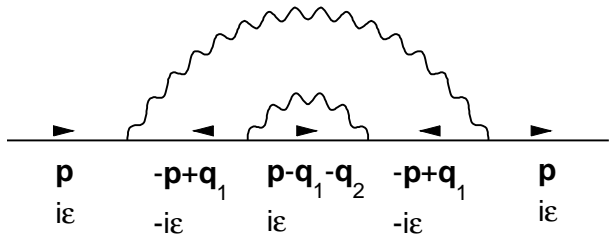
Fig.10. Self-energy diagrams in the model of SC--fluctuations:

(a)---first order diagram with explanation of the meaning of interaction
line (fluctuation propagator of Cooper pairs).
 
(b)---second order diagram.

\vskip 0.3cm

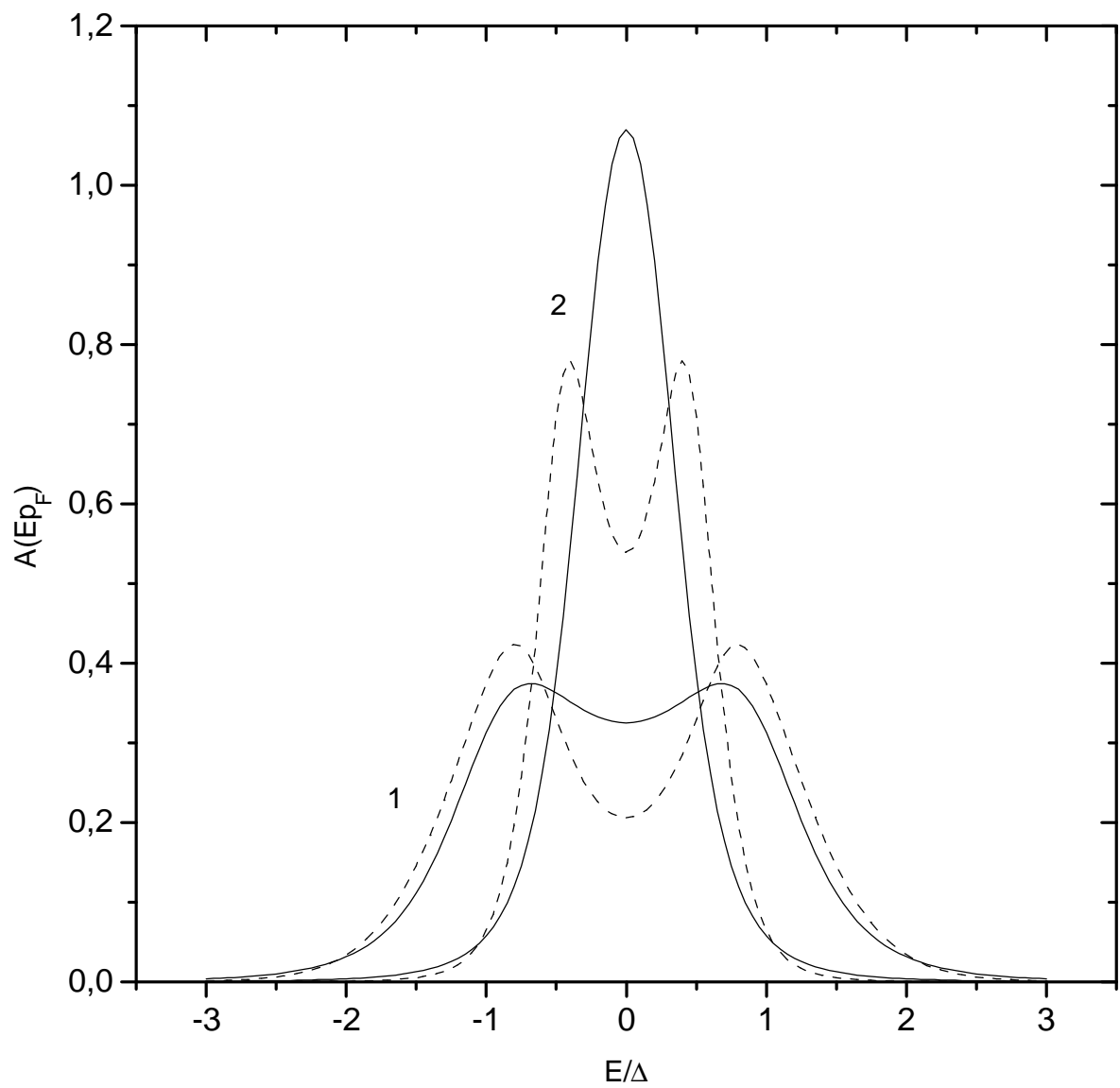
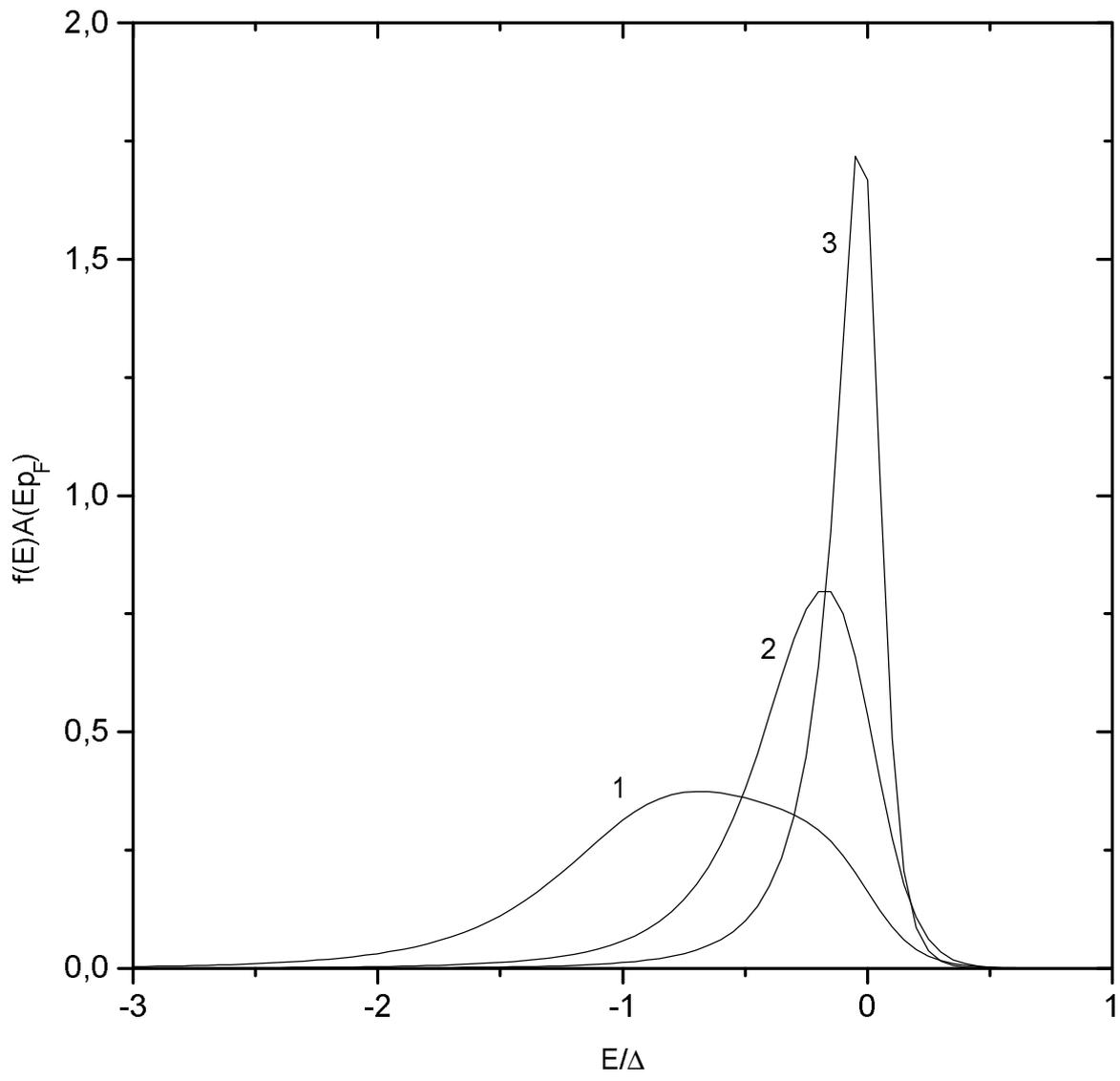
Fig.11.\ (a)--Energy dependence of the spectral density $A(E,{\bf p})$  for
the case of $d$-wave fluctuation pairing for different values of the polar
angle $\phi$, defining the direction of electronic momentum in the plane:

(1)---$\phi=0$;\ (2)---$\phi=\pi/6$.

Correlation length corresponds to $v_F\kappa/\Delta=0.5$ 
(full curve) and $0.1$ (dashed).

(b)--analogous dependence of the product $f(E)A(E,{\bf p})$ ($f(E)$--
Fermi function):

(1)---$\phi=0$;\ (2)---$\phi=\pi/6$;\ (3)---$\phi=\pi/4.83$.

Temperature (in Fermi function)  $T=0.1\Delta$,\ $v_F\kappa/\Delta=0.5$ .

\vskip 0.3cm

Fig.12. Schematic picture of Fermi surface ``destruction'' by pseudogap
due to fluctuational $d$-wave pairing. Dashed are the regions, where the
spectral density is essentially non Fermi-liquid like.

\vskip 0.3cm

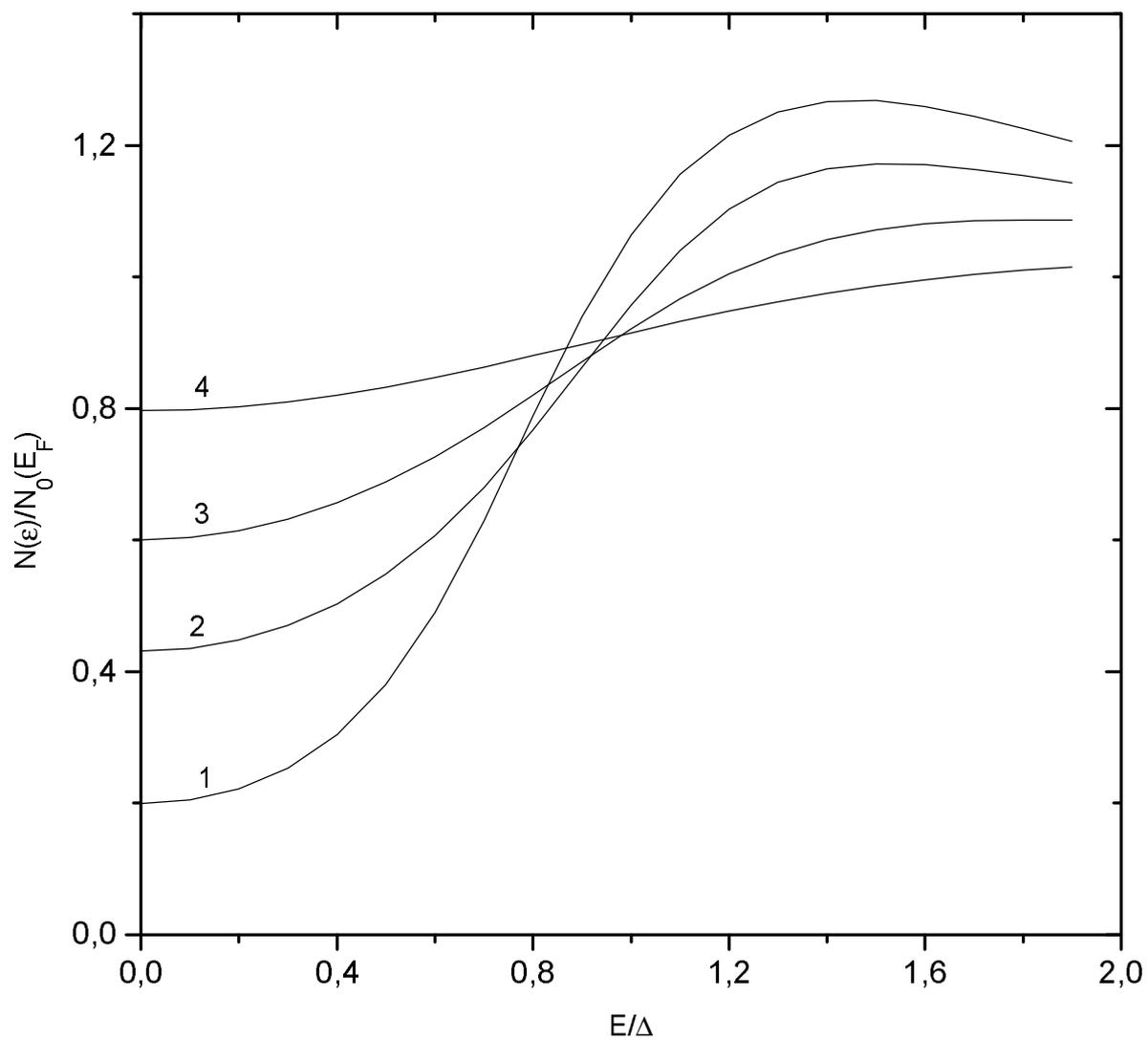
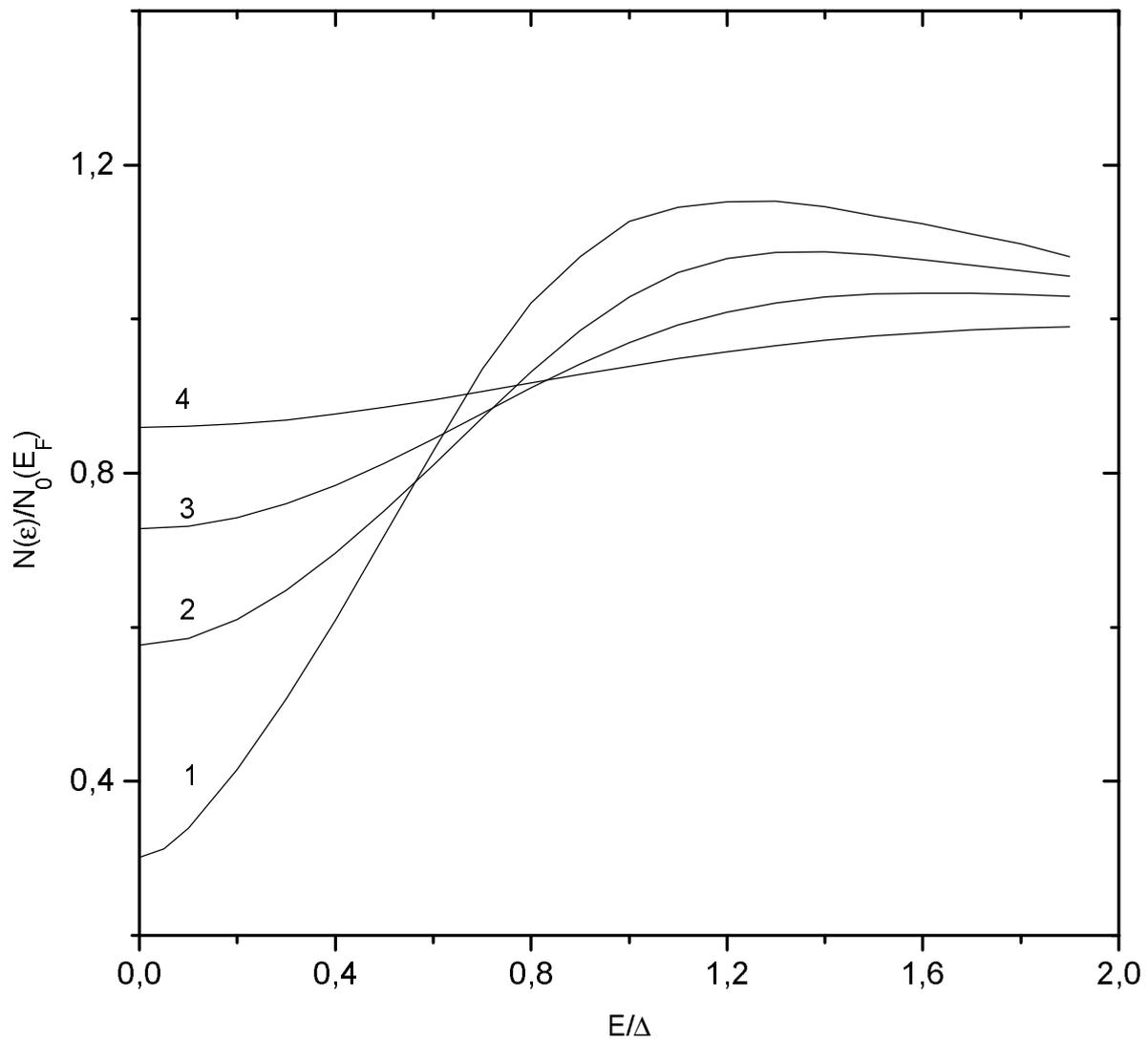
Fig.13. One-electron density of states in the model of SC--fluctuations for 
the different values of parameter $v_F\kappa/\Delta$:

(a)---case of $s$-wave pairing.

(b)---case of $d$-wave pairing.

Curves are shown for the following values of $v_F\kappa/\Delta$:

(1)---0.1;\  (2)---0.5;\  (3)---1.0;\ (4)---2.0.

\vskip 0.3cm

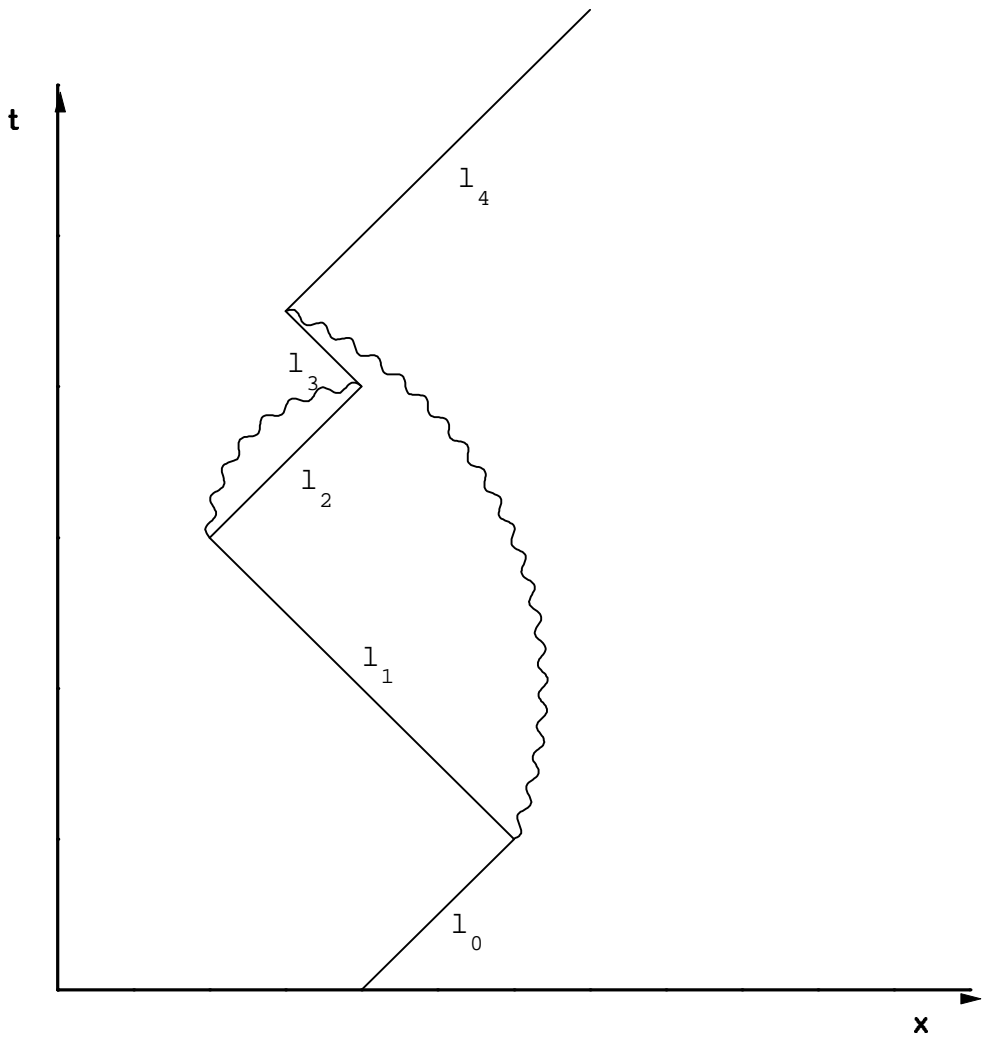
Fig.14. Second order diagram for the correction to the Green's function in
coordinate representation.

\vskip 0.3cm

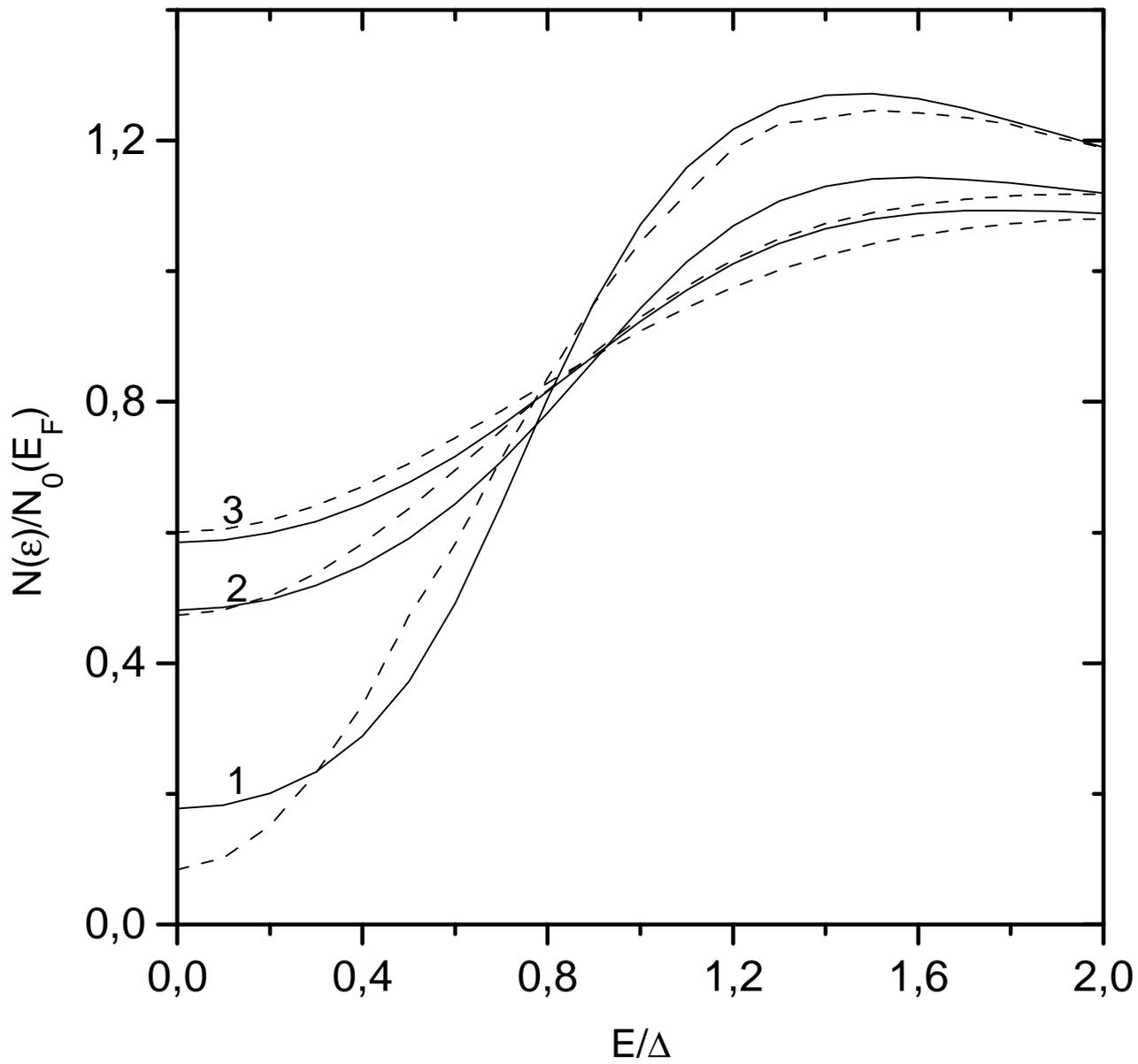
Fig.15. One-electron density of states in one-dimensional model for the
different values of  $v_F\kappa/\Delta$: 

(1)---0.1;\ (2)---0.8;\ (3)---1.2.

Full curves --- result of calculations using (\ref{Ansatz}),(\ref{rec}) 
\cite{C79},\ dashed line --- result of calculations using (\ref{dosalt}).

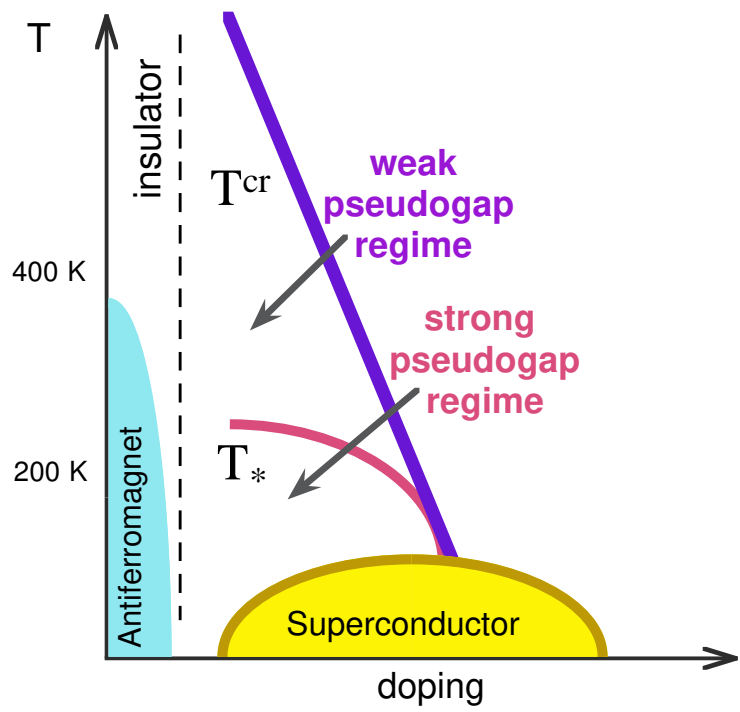
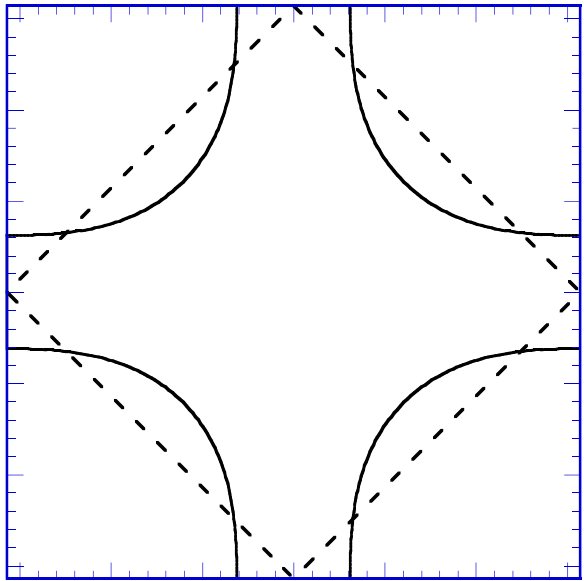
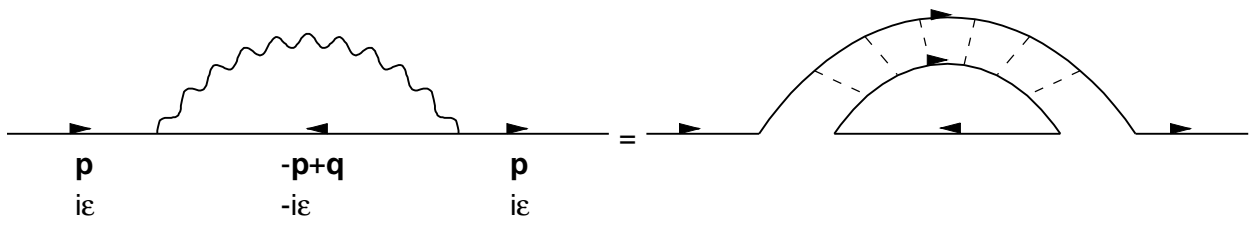
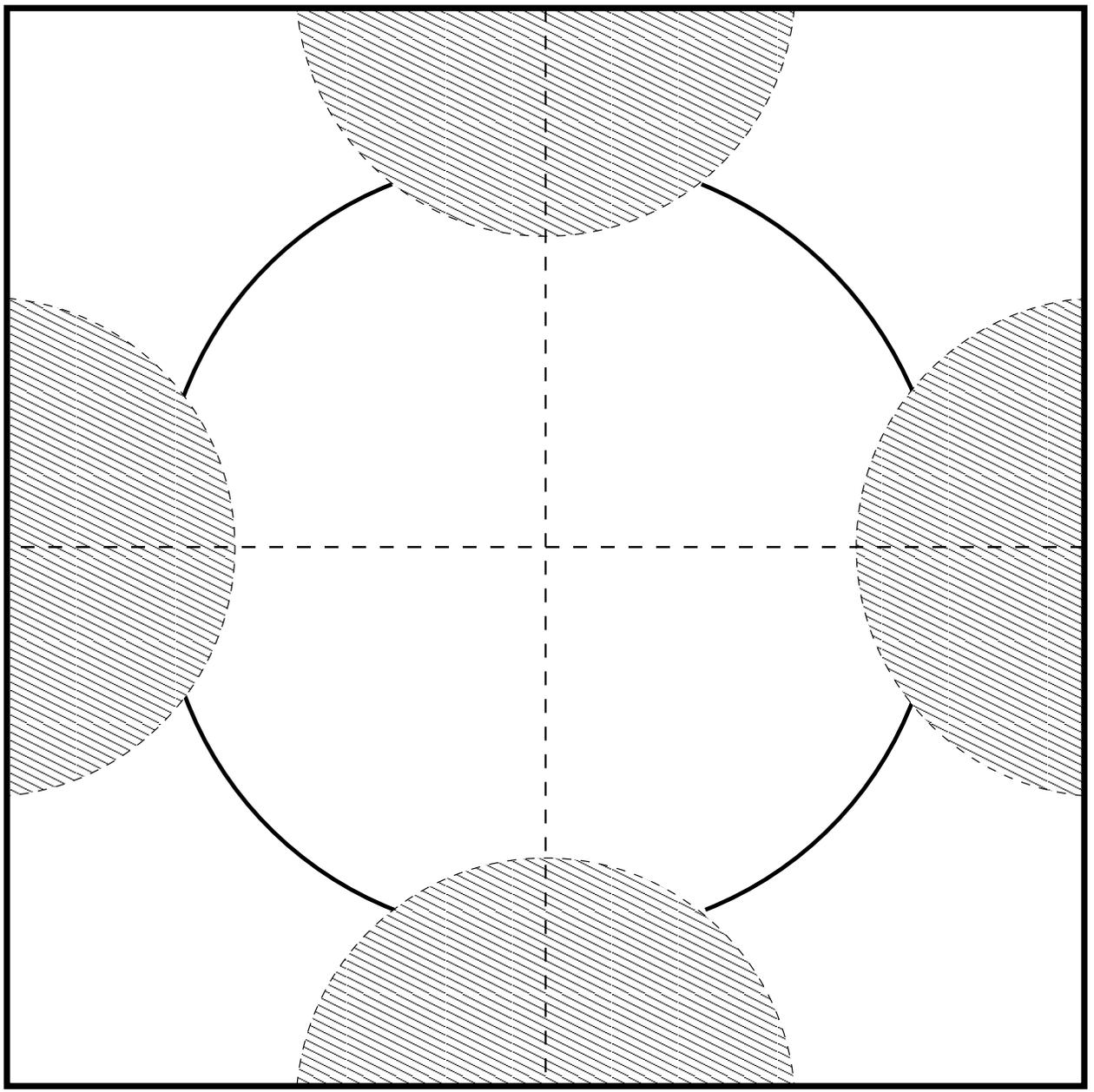
\end{document}